\documentclass{cernyrep} 
\usepackage{texnames}
\usepackage[T1]{fontenc}
\usepackage{graphicx}
\usepackage{subfigure}
\usepackage{pstricks,pstricks-add, multido, pst-coil,pst-solides3d,pst-grad,pst-optic,pst-text}
\usepackage{slantsc}
\usepackage{psfrag}
\newcommand{\qed}{\nobreak \ifvmode \relax \else
      \ifdim\lastskip<1.5em \hskip-\lastskip
      \hskip1.5em plus0em minus0.5em \fi \nobreak
      \vrule height0.75em width0.5em depth0.25em\fi}

\begin{document}
\title{RF engineering basic concepts: the Smith chart}
 
\author{F. Caspers}

\institute{CERN, Geneva, Switzerland}

\maketitle 

\begin{abstract}
The Smith chart is a very valuable and important tool that facilitates interpretation of S-parameter measurements. This paper will give a brief overview on why and more importantly on how to use the chart. Its definition as well as an introduction on how to navigate inside the chart are illustrated. Useful examples show the broad possibilities for use of the chart in a variety of applications. 
\end{abstract}
 
\section{Motivation}

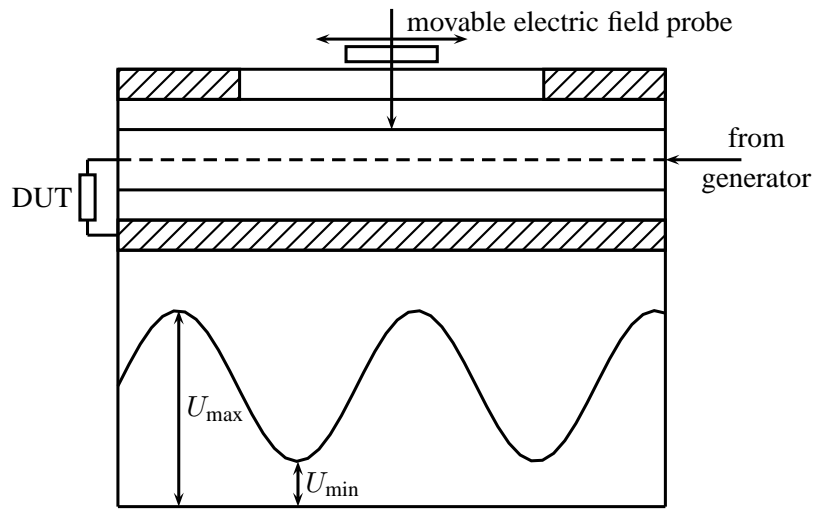
\begin{figure}[htbp]
\centering\begin{pspicture}(0,1)(10,8)%
%
\psset{linewidth=0.04}%
%
\psline(1.4,1.4)(1.4,7.2)%
\psline(8.6,1.4)(8.6,7.2)%
\psline(1.4,1.4)(8.6,1.4)%
%
\pspolygon[fillstyle=hlines](1.4,4.8)(8.6,4.8)(8.6,5.2)(1.4,5.2)%
\psline(1.4,4.8)(8.6,4.8)%
\psline(1.4,5.2)(8.6,5.2)%
\psline(1.4,5.6)(8.6,5.6)%
\psline[linestyle=dashed](1.4,6)(8.6,6)%
\psline(1.4,6.4)(8.6,6.4)%
\pspolygon[fillstyle=hlines](1.4,6.8)(3,6.8)(3,7.2)(1.4,7.2)%
\pspolygon[fillstyle=hlines](7,6.8)(8.6,6.8)(8.6,7.2)(7,7.2)%
\psline(1.4,6.8)(8.6,6.8)%
\psline(1.4,7.2)(8.6,7.2)%
%
%
\psline(1.4,6)(1,6)%
\psline(1,6)(1,5.8)%
\psline(1.4,5)(1,5)%
\psline(1,5)(1,5.2)%
\pspolygon(0.9,5.2)(1.1,5.2)(1.1,5.8)(0.9,5.8)%
\rput(0.4,5.5){DUT}%
\psline[arrows=<-](8.6,6)(9.6,6)%
\rput(9.8,6.3){from}%
\rput(9.8,5.7){generator}%
%
%
\pspolygon(4.4,7.3)(5.6,7.3)(5.6,7.5)(4.4,7.5)%
\psline[arrows=<->](4,7.6)(6,7.6)%
\psline[arrows=->](5,8)(5,6.4)%
\rput[l](5.2,7.8){movable electric field probe}%
%
%
\rput(1.4,3){%
\psplot[algebraic=true]{0}{7.2}{sin(2*x)}%
}%
\psline[arrows=<->](3.768,2)(3.768,1.4)%
\psline[arrows=<->](2.2,4)(2.2,1.4)%
\rput[l](3.868,1.7){$U_{\text{min}}$}%
\rput[l](2.3,2.7){$U_{\text{max}}$}%
\end{pspicture}
\caption{Schematic view of a measurement set--up used to determine the reflection coefficient as well as the voltage standing wave ratio of a device under test (DUT) \cite{meinkegundlach}}
\label{coax}
\end{figure}

With the equipment at hand today, it has become rather easy to measure the reflection factor $\Gamma$ even for complicated networks. In the ``good old days'' though, this was done measuring the electric field strength\footnote{The electrical field strength was used since it can be measured considerably more easily than the magnetic field strength.} at a coaxial measurement line with a slit at different positions in the axial direction (Fig.\,\ref{coax}).
A small electric field probe, protruding into the field region of the coaxial line near the outer conductor, was moved along the line. Its signal was picked up and displayed on a microvoltmeter after rectification via a microwave diode. While moving the probe, field maxima and minima as well as their position and spacing could be found. From this the reflection factor $\Gamma$ and the \textbf{V}oltage \textbf{S}tanding \textbf{W}ave \textbf{R}atio (VSWR or SWR) could be determined using the following definitions: 
\begin{itemize}
	\item $\Gamma$ is defined as the ratio of the electrical field strength $E$ of the reflected wave over the forward travelling wave:
\begin{equation}
	\Gamma = \frac{E\text{ of reflected wave}}{E \text{ of forward traveling wave}}\hspace{0.2cm}.
\label{eq:1}
\end{equation}
	\item The VSWR is defined as the ratio of maximum to minimum measured voltage:
\begin{equation}
	\text{VSWR} = \frac{U_\text{max}}{U_{\text{min}}} = \frac{1 + |\Gamma|}{1 - |\Gamma|}\hspace{0.2cm}.
\label{eq:2}
\end{equation}
\end{itemize}
Although today these measurements are far easier to conduct, the definitions of the aforementioned quantities are still valid. Also their importance has not diminished in the field of microwave engineering and so the reflection coefficient as well as the VSWR are still a vital part of the everyday life of a microwave engineer be it for simulations or measurements. 

A special diagram is widely used to visualize and to facilitate the determination of these quantities. Since it was invented in 1939 by the engineer Phillip Smith, it is simply known as the Smith chart \cite{smith2000}.
\section{Definition of the Smith chart}
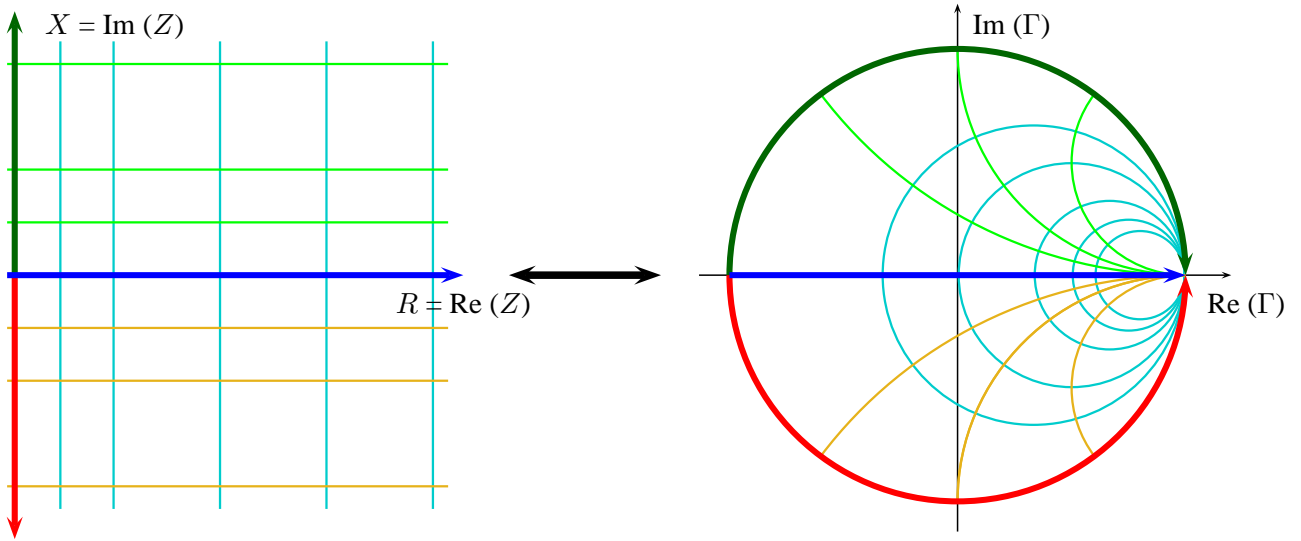
\begin{figure}[htbp]
\centering\begin{pspicture}(0,0)(17,8)%
\psset{linewidth=0.08}
\newrgbcolor{darkgreen}{0.0 0.4 0.0}%
\newrgbcolor{lightblue}{0.0 0.8 0.8}%
\newrgbcolor{darkyellow}{0.9 0.7 0.1}%
\rput(0,-2){%
%
%
\rput(0.5,0){%
\psline[linewidth=0.02,arrows=->](8.6,6)(15.6,6)%
\psline[linewidth=0.02,arrows=->](12,2.6)(12,9.6)%
\pscircle[linecolor=lightblue,linewidth=0.03](14.4,6){0.6}%
\pscircle[linecolor=lightblue,linewidth=0.03](14.25,6){0.75}%
\pscircle[linecolor=lightblue,linewidth=0.03](14,6){1}%
\pscircle[linecolor=lightblue,linewidth=0.03](13.5,6){1.5}%
\pscircle[linecolor=lightblue,linewidth=0.03](13,6){2}%
\psarcAB[linecolor=green,linewidth=0.03](15,9)(12,9)(15,6)%
\psarcAB[linecolor=green,linewidth=0.03](15,7.5)(13.8,8.4)(15,6)%
\psarcAB[linecolor=green,linewidth=0.03](15,12)(10.2,8.4)(15,6)%
\psarcAB[linecolor=darkyellow,linewidth=0.03](15,3)(15,6)(12,3)%
\psarc[linecolor=darkyellow,linewidth=0.03](15,3){3}{90}{180}%
\psarcAB[linecolor=darkyellow,linewidth=0.03](15,4.5)(15,6)(13.8,3.6)%
\psarcAB[linecolor=darkyellow,linewidth=0.03](15,0)(15,6)(10.2,3.6)%
\psarc[linecolor=darkgreen,arrows=<-](12,6){3}{0}{180}%
\psarc[linecolor=red,arrows=->](12,6){3}{180}{0}%
\psline[linecolor=blue,arrows=->](9,6)(15,6)%
\rput[l](12.2,9.3){Im ($\Gamma$)}
\rput(15.8,5.6){Re ($\Gamma$)}
}%
%
%
\psline[linecolor=lightblue,linewidth=0.03](0.7,2.9)(0.7,9.1)%
\psline[linecolor=lightblue,linewidth=0.03](1.4,2.9)(1.4,9.1)%
\psline[linecolor=lightblue,linewidth=0.03](2.8,2.9)(2.8,9.1)%
\psline[linecolor=lightblue,linewidth=0.03](4.2,2.9)(4.2,9.1)%
\psline[linecolor=lightblue,linewidth=0.03](5.6,2.9)(5.6,9.1)%
\psline[linecolor=green,linewidth=0.03](0,6.7)(5.8,6.7)%
\psline[linecolor=green,linewidth=0.03](0,7.4)(5.8,7.4)%
\psline[linecolor=green,linewidth=0.03](0,8.8)(5.8,8.8)%
\psline[linecolor=darkyellow,linewidth=0.03](0,5.3)(5.8,5.3)%
\psline[linecolor=darkyellow,linewidth=0.03](0,4.6)(5.8,4.6)%
\psline[linecolor=darkyellow,linewidth=0.03](0,3.2)(5.8,3.2)%
\psline[linecolor=darkgreen,arrows=->](0.1,6)(0.1,9.5)%
\psline[linecolor=red,arrows=->](0.1,6)(0.1,2.5)%
\psline[linecolor=blue,arrows=->](0,6)(6,6)%
\rput[l](0.5,9.3){$X$ = Im ($Z$)}%
\rput(6,5.6){$R$ = Re ($Z$)}%
%
%
\psline[arrows=<->,linewidth=0.1](6.6,6)(8.6,6)%
%
%
}%
\end{pspicture}%
\caption{Illustration of the Moebius transform from the complex impedance plane to the $\Gamma$ plane commonly known as Smith chart}
\label{scbasic}
\end{figure}

The Smith chart provides a graphical representation of $\Gamma$ that permits the determination of quantities such as the VSWR or the terminating impedance of a device under test (DUT). It uses a bilinear Moebius transformation, projecting the complex impedance plane onto the complex $\Gamma$ plane:
\begin{equation}
	\Gamma = \frac{Z - Z_{\text{0}}}{Z + Z_{\text{0}}} \hspace{0.5cm}\text{ with }\hspace{0.5cm} Z = R + \text{j}\,X\hspace{0.2cm}.
\label{eq:3}
\end{equation}
As can be seen in Fig.\,\ref{scbasic} the half-plane with positive real part of impedance $Z$ is mapped onto the interior of the unit circle of the $\Gamma$ plane. For a detailed calculation see Appendix \ref{appendix1}.
\subsection{Properties of the transformation}
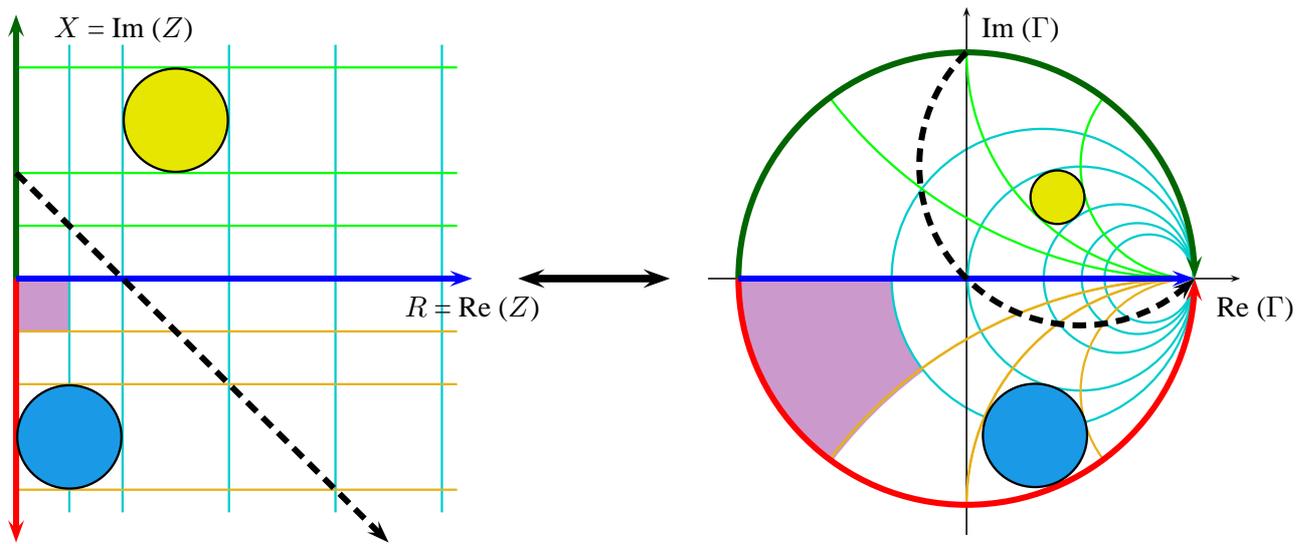
\begin{figure}[htbp]
\centering\begin{pspicture}(0,0)(17,8)%
\psset{linewidth=0.08}
\newrgbcolor{darkgreen}{0.0 0.4 0.0}%
\newrgbcolor{lightblue}{0.0 0.8 0.8}%
\newrgbcolor{darkyellow}{0.9 0.7 0.1}%
\newrgbcolor{yellow}{0.9 0.9 0.0}%
\newrgbcolor{middleblue}{0.1 0.6 0.9}%
\newrgbcolor{lilac}{0.8 0.6 0.8}%
\rput(0,-2){%
%
%
%
\psarc[linestyle=none,fillcolor=lilac,fillstyle=solid](12.5,6){3}{180}{0}%
\rput(0.5,0){%
\pscircle[linestyle=none,fillcolor=white,fillstyle=solid](15,0){6}%
\pscircle[linecolor=lightblue,linewidth=0.03,fillcolor=white,fillstyle=solid](13,6){2}%
\psline[linewidth=0.02,arrows=->](8.6,6)(15.6,6)%
\psline[linewidth=0.02,arrows=->](12,2.6)(12,9.6)%
\pscircle[linecolor=lightblue,linewidth=0.03](14.4,6){0.6}%
\pscircle[linecolor=lightblue,linewidth=0.03](14.25,6){0.75}%
\pscircle[linecolor=lightblue,linewidth=0.03](14,6){1}%
\pscircle[linecolor=lightblue,linewidth=0.03](13.5,6){1.5}%
\psarcAB[linecolor=green,linewidth=0.03](15,9)(12,9)(15,6)%
\psarcAB[linecolor=green,linewidth=0.03](15,7.5)(13.8,8.4)(15,6)%
\psarcAB[linecolor=green,linewidth=0.03](15,12)(10.2,8.4)(15,6)%
\psarcAB[linecolor=darkyellow,linewidth=0.03](15,3)(15,6)(12,3)%
\psarcAB[linecolor=darkyellow,linewidth=0.03](15,0)(15,6)(10.2,3.6)%
\psarc[linecolor=darkyellow,linewidth=0.03](15,3){3}{90}{180}%
\psarcAB[linecolor=darkyellow,linewidth=0.03](15,4.5)(15,6)(13.8,3.6)%
\psarcAB[linecolor=darkyellow,linewidth=0.03](15,0)(15,6)(10.2,3.6)%
\psarc[linecolor=darkgreen,arrows=<-](12,6){3}{0}{180}%
\psarc[linecolor=red,arrows=->](12,6){3}{180}{0}%
\psline[linecolor=blue,arrows=->](9,6)(15,6)%
\rput[l](12.2,9.3){Im ($\Gamma$)}
\rput(15.8,5.6){Re ($\Gamma$)}
%
%
\psarcAB[linestyle=dashed,arrows=->](13.5,7.5)(12,9)(15,6)%
\pscircle[linewidth=0.03,fillcolor=yellow,fillstyle=solid](13.1954,7.07882){0.37}%
\pscircle[linewidth=0.03,fillcolor=middleblue,fillstyle=solid](12.9,3.92){0.7}%
}%
%
%
%
\pspolygon[fillcolor=lilac,fillstyle=solid,linecolor=lilac,linewidth=0.03](0,6)(0.7,6)(0.7,5.3)(0,5.3)%
%
\psline[linecolor=lightblue,linewidth=0.03](0.7,2.9)(0.7,9.1)%
\psline[linecolor=lightblue,linewidth=0.03](1.4,2.9)(1.4,9.1)%
\psline[linecolor=lightblue,linewidth=0.03](2.8,2.9)(2.8,9.1)%
\psline[linecolor=lightblue,linewidth=0.03](4.2,2.9)(4.2,9.1)%
\psline[linecolor=lightblue,linewidth=0.03](5.6,2.9)(5.6,9.1)%
\psline[linecolor=green,linewidth=0.03](0,6.7)(5.8,6.7)%
\psline[linecolor=green,linewidth=0.03](0,7.4)(5.8,7.4)%
\psline[linecolor=green,linewidth=0.03](0,8.8)(5.8,8.8)%
\psline[linecolor=darkyellow,linewidth=0.03](0,5.3)(5.8,5.3)%
\psline[linecolor=darkyellow,linewidth=0.03](0,4.6)(5.8,4.6)%
\psline[linecolor=darkyellow,linewidth=0.03](0,3.2)(5.8,3.2)%
\psline[linecolor=darkgreen,arrows=->](0,6)(0,9.5)%
\psline[linecolor=red,arrows=->](0,6)(0,2.5)%
\psline[linecolor=blue,arrows=->](0,6)(6,6)%
%
%
\pscircle[fillcolor=yellow,fillstyle=solid,linewidth=0.03](2.1,8.1){0.7}%
\pscircle[fillcolor=middleblue,fillstyle=solid,linewidth=0.03](0.7,3.9){0.7}%
\psline[linestyle=dashed, arrows=->](0,7.4)(4.9,2.5)%

%
\rput[l](0.5,9.3){$X$ = Im ($Z$)}%
\rput(6,5.6){$R$ = Re ($Z$)}%
%
%
\psline[arrows=<->,linewidth=0.1](6.6,6)(8.6,6)%
}%
\end{pspicture}%
\caption{Illustration of the transformation of basic shapes from the $Z$ to the $\Gamma$ plane}
\label{prop}
\end{figure}

In general, this transformation has two main properties:
\begin{itemize}
	\item generalized circles are transformed into generalized circles (note that a straight line is nothing else than a circle with infinite radius and is therefore mapped as a circle in the Smith chart);
	\item angles are preserved locally.
\end{itemize}
Figure \ref{prop} illustrates how certain basic shapes transform from the impedance to the $\Gamma$ plane.
\subsection{Normalization}
The Smith chart is usually normalized to a terminating impedance $Z_{\text{0}}$ (= real):
\begin{equation}
	z = \frac{Z}{Z_{\text{0}}}\hspace{0.2cm}.
\label{eq:4}
\end{equation}
This leads to a simplification of the transform:
\begin{equation}
	\Gamma = \frac{z - 1}{z + 1} \hspace{0.5cm} \Leftrightarrow \hspace{0.5cm} z = \frac{1 + \Gamma}{1 - \Gamma}\hspace{0.2cm}.
\label{eq:5}
\end{equation}
Although $Z$ = 50\,$\Omega$ is the most common reference impedance (characteristic impedance of coaxial cables) and many applications use this normalization, any other real and positive value is possible. \textit{Therefore it is crucial to check the normalization before using any chart.}

Commonly used charts that map the impedance plane onto the $\Gamma$ plane always look confusing at first, as many circles are depicted (Fig.\,\ref{smith}). Keep in mind that all of them can be calculated as shown in Appendix \ref{appendix1} and that this representation is the same as shown in all previous figures --- it just contains more circles.
\begin{figure}[htbp]
\centering\includegraphics[width=.9\linewidth]{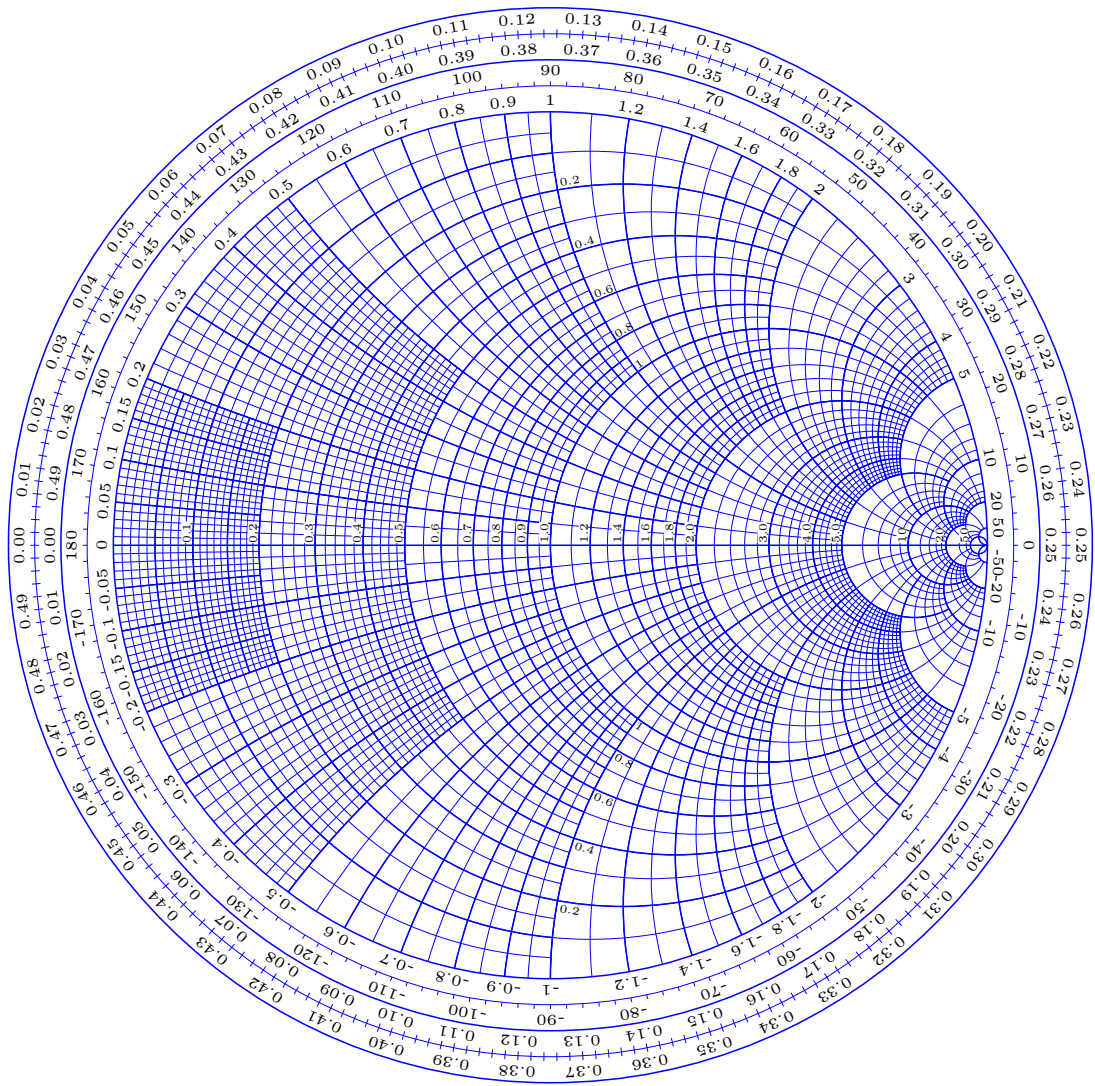}
\caption{Example of a commonly used Smith chart}
\label{smith}
\end{figure}

\subsection{Admittance plane}
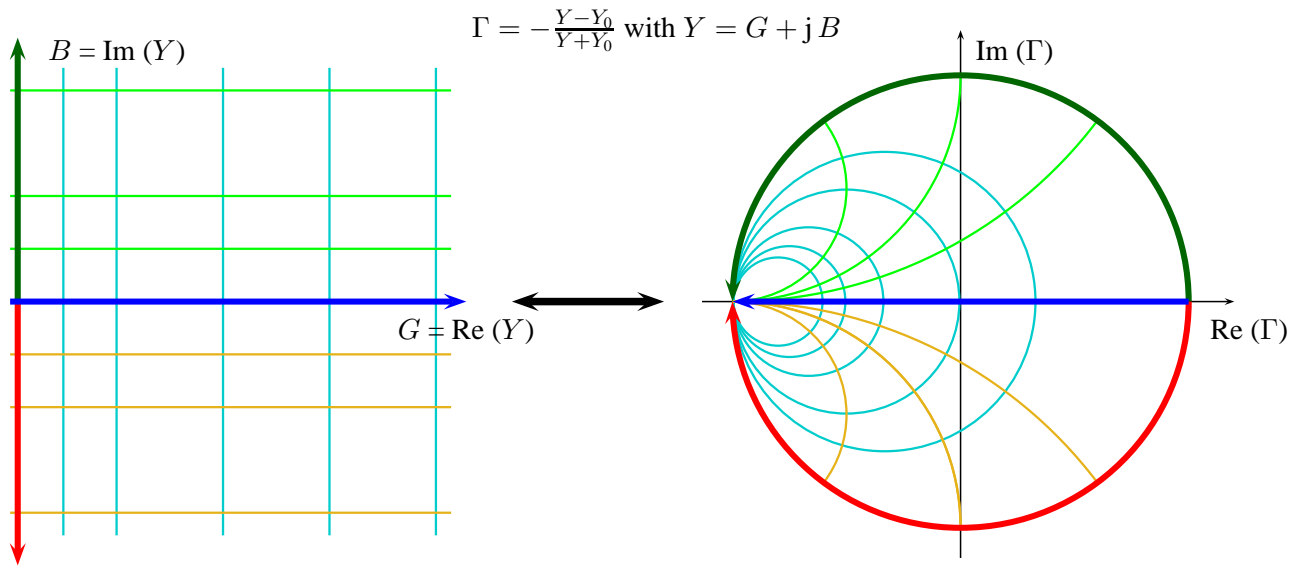
\begin{figure}[htbp]
\centering\begin{pspicture}(0,0)(17,8)%
\psset{linewidth=0.08}
\newrgbcolor{darkgreen}{0.0 0.4 0.0}%
\newrgbcolor{lightblue}{0.0 0.8 0.8}%
\newrgbcolor{darkyellow}{0.9 0.7 0.1}%
\rput(0,-2){%
%
%
\rput(0.5,0){%
\psline[linewidth=0.02,arrows=->](8.6,6)(15.6,6)%
\psline[linewidth=0.02,arrows=->](12,2.6)(12,9.6)%
\symPlan(12,0)(12,12){%
\pscircle[linecolor=lightblue,linewidth=0.03](14.4,6){0.6}%
\pscircle[linecolor=lightblue,linewidth=0.03](14.25,6){0.75}%
\pscircle[linecolor=lightblue,linewidth=0.03](14,6){1}%
\pscircle[linecolor=lightblue,linewidth=0.03](13.5,6){1.5}%
\pscircle[linecolor=lightblue,linewidth=0.03](13,6){2}%
\psarcAB[linecolor=green,linewidth=0.03](15,9)(12,9)(15,6)%
\psarcAB[linecolor=green,linewidth=0.03](15,7.5)(13.8,8.4)(15,6)%
\psarcAB[linecolor=green,linewidth=0.03](15,12)(10.2,8.4)(15,6)%
\psarcAB[linecolor=darkyellow,linewidth=0.03](15,3)(15,6)(12,3)%
\psarc[linecolor=darkyellow,linewidth=0.03](15,3){3}{90}{180}%
\psarcAB[linecolor=darkyellow,linewidth=0.03](15,4.5)(15,6)(13.8,3.6)%
\psarcAB[linecolor=darkyellow,linewidth=0.03](15,0)(15,6)(10.2,3.6)%
\psarc[linecolor=darkgreen,arrows=<-](12,6){3}{0}{180}%
\psarc[linecolor=red,arrows=->](12,6){3}{180}{0}%
\psline[linecolor=blue,arrows=->](9,6)(15,6)%
}%
\rput[l](12.2,9.3){Im ($\Gamma$)}
\rput(15.8,5.6){Re ($\Gamma$)}
}%
%
%
%
\psline[linecolor=lightblue,linewidth=0.03](0.7,2.9)(0.7,9.1)%
\psline[linecolor=lightblue,linewidth=0.03](1.4,2.9)(1.4,9.1)%
\psline[linecolor=lightblue,linewidth=0.03](2.8,2.9)(2.8,9.1)%
\psline[linecolor=lightblue,linewidth=0.03](4.2,2.9)(4.2,9.1)%
\psline[linecolor=lightblue,linewidth=0.03](5.6,2.9)(5.6,9.1)%
\psline[linecolor=green,linewidth=0.03](0,6.7)(5.8,6.7)%
\psline[linecolor=green,linewidth=0.03](0,7.4)(5.8,7.4)%
\psline[linecolor=green,linewidth=0.03](0,8.8)(5.8,8.8)%
\psline[linecolor=darkyellow,linewidth=0.03](0,5.3)(5.8,5.3)%
\psline[linecolor=darkyellow,linewidth=0.03](0,4.6)(5.8,4.6)%
\psline[linecolor=darkyellow,linewidth=0.03](0,3.2)(5.8,3.2)%
\psline[linecolor=darkgreen,arrows=->](0.1,6)(0.1,9.5)%
\psline[linecolor=red,arrows=->](0.1,6)(0.1,2.5)%
\psline[linecolor=blue,arrows=->](0,6)(6,6)%
\rput[l](0.5,9.3){$B$ = Im ($Y$)}%
\rput(6,5.6){$G$ = Re ($Y$)}%
%
%
\psline[arrows=<->,linewidth=0.1](6.6,6)(8.6,6)%
%
%
\rput(8.5,9.6){$\Gamma = -\frac{Y - Y_{\text{0}}}{Y + Y_{\text{0}}} \text{ with } Y = G + \text{j}\,B$}%
}%
\end{pspicture}%
\caption{Mapping of the admittance plane into the $\Gamma$ plane}
\label{admit}
\end{figure}

The Moebius transform that generates the Smith chart provides also a mapping of the complex admittance plane ($Y = \frac{1}{Z}$ or normalized $y = \frac{1}{z}$) into the same chart:
\begin{equation}
	\Gamma = -\frac{y - \text{1}}{y + \text{1}} = -\frac{Y - Y_{\text{0}}}{Y + Y_{\text{0}}} = - \frac{1/Z - 1/Z_{\text{0}}}{1/Z + 1/Z_{\text{0}}} = \frac{Z - Z_{\text{0}}}{Z + Z_{\text{0}}} = \frac{z - \text{1}}{z + \text{1}}\hspace{0.2cm}.
\label{eq:6}
\end{equation}
Using this transformation, the result is the same chart, but mirrored at the centre of the Smith chart (Fig.\,\ref{admit}).
Often both mappings, the admittance and the impedance plane, are combined into one chart, which looks even more confusing (see last page). For reasons of simplicity all illustrations in this paper will use only the mapping from the impedance to the $\Gamma$ plane.
\section{Navigation in the Smith chart}
The representation of circuit elements in the Smith chart is discussed in this section starting with the important points inside the chart. Then several examples of circuit elements will be given and their representation in the chart will be illustrated.
\subsection{Important points}
\begin{figure}[htbp]
\centering\begin{pspicture}(0,0)(10,8)%
\newrgbcolor{darkgreen}{0.0 0.4 0.0}%
\newrgbcolor{lightblue}{0.0 0.8 0.8}%
\newrgbcolor{darkyellow}{0.9 0.7 0.1}%
%
%
\rput(-7,-2){%
\psline[linewidth=0.02,arrows=->](8.6,6)(15.6,6)%
\psline[linewidth=0.02,arrows=->](12,2.6)(12,9.6)%
\pscircle[linecolor=lightblue,linewidth=0.03](14.4,6){0.6}%
\pscircle[linecolor=lightblue,linewidth=0.03](14.25,6){0.75}%
\pscircle[linecolor=lightblue,linewidth=0.03](14,6){1}%
\pscircle[linecolor=lightblue,linewidth=0.03](13.5,6){1.5}%
\pscircle[linecolor=lightblue,linewidth=0.03](13,6){2}%
\psarcAB[linecolor=green,linewidth=0.03](15,9)(12,9)(15,6)%
\psarcAB[linecolor=green,linewidth=0.03](15,7.5)(13.8,8.4)(15,6)%
\psarcAB[linecolor=green,linewidth=0.03](15,12)(10.2,8.4)(15,6)%
\psarcAB[linecolor=darkyellow,linewidth=0.03](15,3)(15,6)(12,3)%
\psarc[linecolor=darkyellow,linewidth=0.03](15,3){3}{90}{180}%
\psarcAB[linecolor=darkyellow,linewidth=0.03](15,4.5)(15,6)(13.8,3.6)%
\psarcAB[linecolor=darkyellow,linewidth=0.03](15,0)(15,6)(10.2,3.6)%
\psarc[linecolor=darkgreen,arrows=<-,linewidth=0.08](12,6){3}{0}{180}%
\psarc[linecolor=red,arrows=->,linewidth=0.08](12,6){3}{180}{0}%
\psline[linecolor=blue,arrows=->,linewidth=0.08](9,6)(15,6)%
\rput[l](12.2,9.3){Im ($\Gamma$)}
\rput(15.8,5.6){Re ($\Gamma$)}
}%
%
%
\pscircle[fillstyle=solid,fillcolor=black](2,4){0.1}%
\pscircle[fillstyle=solid,fillcolor=black](5,4){0.1}%
\pscircle[fillstyle=solid,fillcolor=black](8,4){0.1}%
\rput[l](7.6,1.4){matched load}%
\rput[r](2,6){short circuit}%
\rput[l](8,6){open circuit}%
\psline[arrows=->,linewidth=0.05](9,5.8)(8.1,4.1)%
\psline[arrows=->,linewidth=0.05](1,5.8)(1.9,4.1)%
\psline[arrows=->,linewidth=0.05](8.6,1.6)(5.1,3.9)%
\end{pspicture}%
\caption{Important points in the Smith chart}
\label{points}
\end{figure}
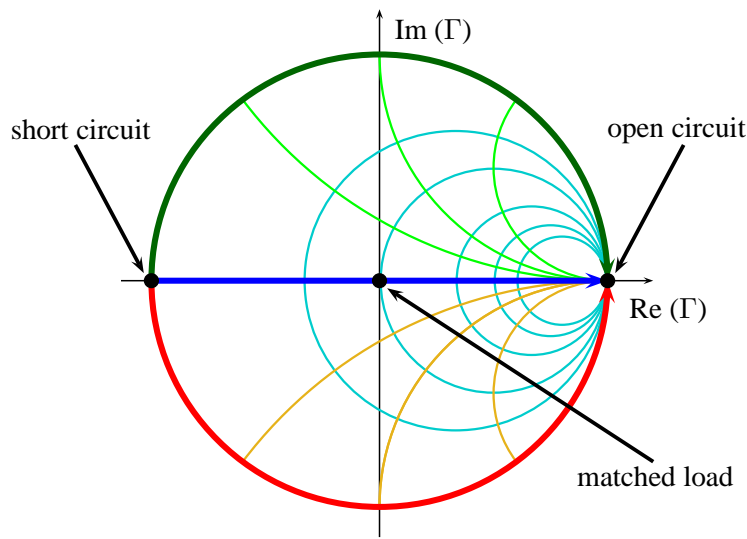

There are three important points in the chart:
\begin{enumerate}
	\item Open circuit with $\Gamma = 1, z \rightarrow \infty$
	\item Short circuit with $\Gamma = -1, z = 0$
	\item Matched load with $\Gamma = 0, z = 1$
\end{enumerate}
They all are located on the real axis at the beginning, the end, and the centre of the circle (Fig.\,\ref{points}).
The upper half of the chart is inductive, since it corresponds to the positive imaginary part of the impedance. The lower half is capacitive as it is corresponding to the negative imaginary part of the impedance.

\begin{figure}[htbp]
\centering\input{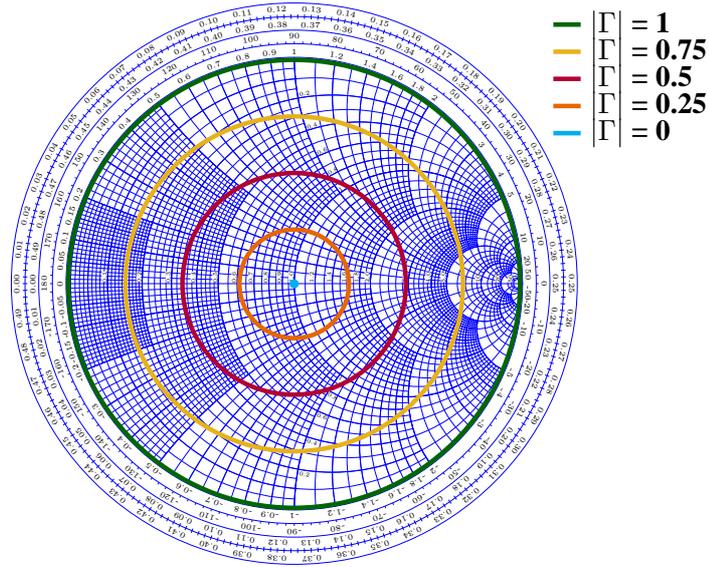}
\caption{Illustration of circles representing a constant reflection factor}
\label{concentric}
\end{figure}

Concentric circles around the diagram centre represent constant reflection factors (Fig.\,\ref{concentric}). Their radius is directly proportional to the magnitude of $\Gamma$, therefore a radius of 0.5 corresponds to reflection of 3 dB (half of the signal is reflected) whereas the outermost circle (radius = 1) represents full reflection.
Therefore matching problems are easily visualized in the Smith chart since a mismatch will lead to a reflection coefficient larger than 0 (Eq.\,(\ref{eq:7})).
\begin{equation}
	\text{Power into the load = forward power - reflected power: }P = \frac{1}{2}\left(\left|a\right|^{2} - \left|b\right|^{2}\right) = \frac{\left|a\right|^{2}}{2}\left(1 - \left|\Gamma\right|^{2}\right)\hspace{0.2cm}.
\label{eq:7}
\end{equation}
In Eq.\,(\ref{eq:7}) the European notation\footnote{The commonly used notation in the US is power = $\left|a\right|^{2}$. These conventions have no impact on S parameters but they are relevant for absolute power calculation. Since this is rarely used in the Smith chart, the definition used is not critical for this paper.} is used, where power = $\frac{\left|a\right|^{2}}{2}$. Furthermore $(1 - \left|\Gamma\right|^{2})$ corresponds to the mismatch loss.

\begin{figure}[htbp]
\centering\input{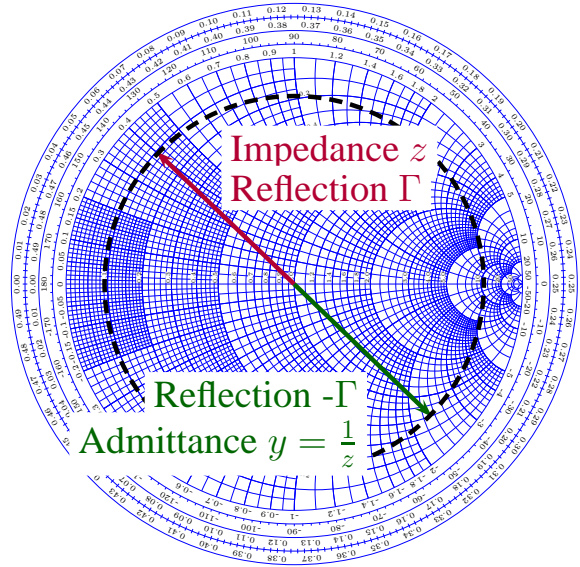}
\caption{Conversion of an impedance to the corresponding emittance in the Smith chart}
\label{imptoad}
\end{figure}

Although only the mapping of the impedance plane to the $\Gamma$ plane is used, one can easily use it to determine the admittance since
\begin{equation}
	\Gamma(\frac{1}{z}) = \frac{\frac{1}{z} - 1}{\frac{1}{z} + 1} = \frac{1 - z}{1 + z} = \left(\frac{z - 1}{z + 1}\right)\text{ or } \Gamma(\frac{1}{z}) = - \Gamma(z)\hspace{0.2cm}.
\label{eq:8}
\end{equation}
In the chart this can be visualized by rotating the vector of a certain impedance by 180$^{\circ}$ (Fig.\,\ref{imptoad}).
\subsection{Adding impedances in series and parallel (shunt)}
\begin{figure}[htbp]
\centering\input{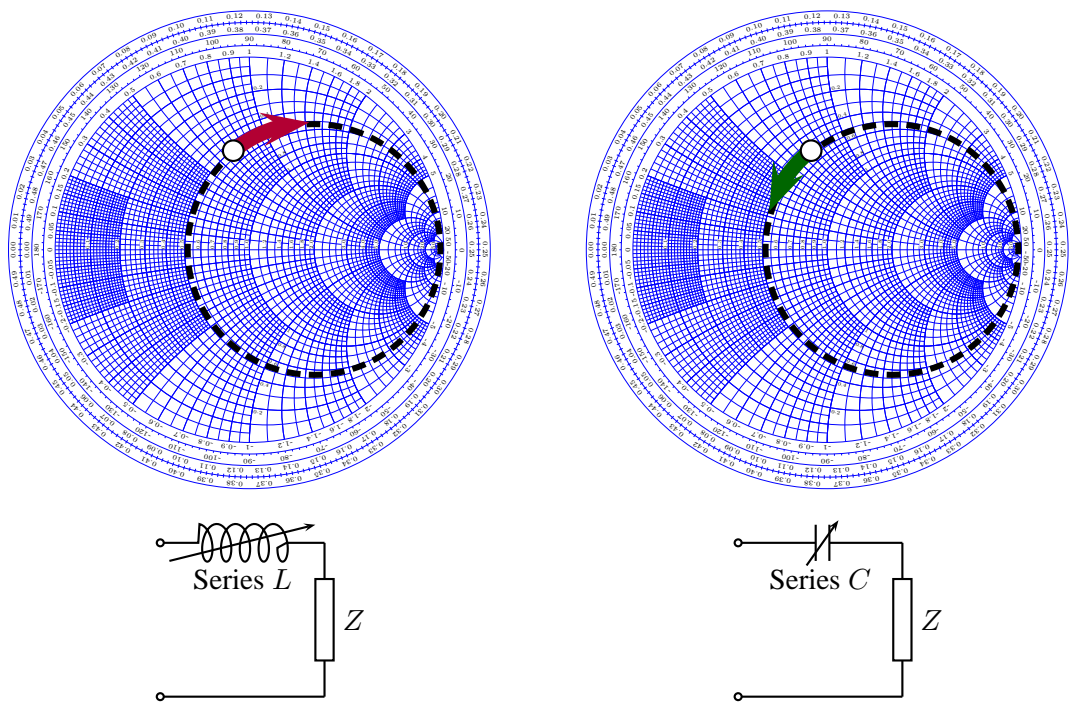}
\caption{Traces of circuits with variable impedances connected in series}
\label{series}
\end{figure}

\begin{figure}[htbp]
\centering\input{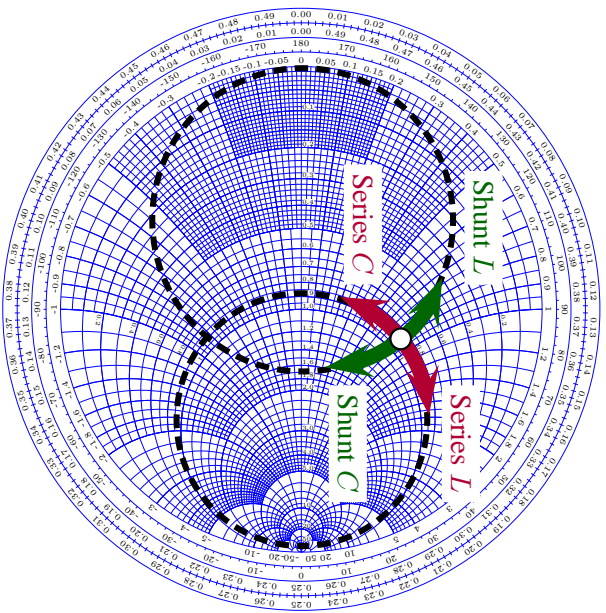}
\caption{Illustration of navigation in the Smith chart when adding lumped elements}
\label{rule}
\end{figure}

A lumped element with variable impedance connected in series is an example of a simple circuit. The corresponding signature of such a circuit for a variable inductance and a variable capacitor is a circle. Depending on the type of impedance, this circle is passed through clockwise (inductance) or anticlockwise (Fig.\,\ref{series}).
If a lumped element is added in parallel, the situation is the same as for an element connected in series mirrored by 180$^{\circ}$ (Fig.\,\ref{para}). This corresponds to taking the same points in the admittance mapping.
\begin{figure}[htbp]
\centering\input{parallel.tex}
\caption{Traces of circuits with variable impedances connected in parallel}
\label{para}
\end{figure}
Summarizing both cases, one ends up with a simple rule for navigation in the Smith chart: \\
\\
%
\textit{For elements connected in series use the circles in the impedance plane. Go clockwise for an added inductance and anticlockwise for an added capacitor. For elements in parallel use the circles in the admittance plane. Go clockwise for an added capacitor and anticlockwise for an added inductance.}\\
\\
%
This rule can be illustrated as shown in Fig.\,\ref{rule}
\subsection{Impedance transformation by transmission line}
The S matrix of an ideal, lossless transmission line of length $l$ is given by
\begin{equation}
	S = \left[ 
\begin{array}{cc}
	0 & \text{e}^{-\text{j}\beta l} \\
	\text{e}^{-\text{j}\beta l} & 0 \\
\end{array}
\right]
\label{eq:9}
\end{equation}
where $\beta = \frac{2\pi}{\lambda}$ is the propagation coefficient with the wavelength $\lambda$ ($\lambda = \lambda_{\text{0}}$ for $\epsilon_{\text{r}} = 1$). 

\begin{figure}[htbp]
\centering\input{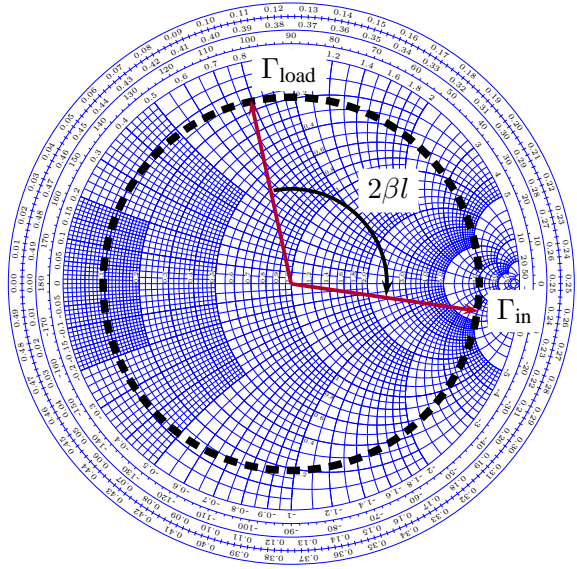}
\caption{Illustration of adding a transmission line of length $l$ to an impedance}
\label{transmissionline}
\end{figure}

When adding a piece of coaxial line, we turn clockwise on the corresponding circle leading to a transformation of the reflection factor $\Gamma _{\text{load}}$ (without line) to the new reflection factor $\Gamma _{\text{in}} = \Gamma _{\text{load}}\text{e}^{-\text{j}2\beta l}$. Graphically speaking, this means that the vector corresponding to $\Gamma_{\text{in}}$ is rotated clockwise by an angle of 2$\beta l$ (Fig.\,\ref{transmissionline}).

\begin{figure}[htbp]
\centering\begin{pspicture}(0,0)(6.56,6.4)%
\newrgbcolor{darkgreen}{0.0 0.4 0.0}%
\psset{unit=0.8cm}
\psset{linewidth=0.05}
\pspolygon[linestyle=none,fillstyle=solid,fillcolor=darkgreen](0,4)(4,4)(4,8)%
\pspolygon[linestyle=none,fillstyle=hlines](0,4)(4,4)(4,8)%
\pspolygon[linestyle=none,fillstyle=solid,fillcolor=red](4,0)(4,4)(8,4)%
\pspolygon[linestyle=none,fillstyle=hlines](4,0)(4,4)(8,4)%
\psline[linecolor=white,linewidth=0.2](8,4)(4,0)%
\psline[linecolor=white,linewidth=0.2](4,8)(0,4)%
\rput(0.1,0){%
\psplot[algebraic=true,fillstyle=solid]{0}{3.8}{tan(0.35*x)+4}%
}%
\rput(-1.1,0){%
\psplot[algebraic=true,fillstyle=solid]{5.2}{8.975979}{tan(0.35*x)+4}%
}%
\psline[linestyle=dashed,linewidth=0.03](4,0)(4,8)%
\psline[arrows=->](0,4)(8.2,4)%
\psline[arrows=->](0,0)(0,8)%
\pscircle[fillstyle=solid,fillcolor=black](4,4){0.1}%
\pscircle[fillstyle=solid,fillcolor=black](8,4){0.1}%
\rput[l](0.4,7.8){Im ($Z$)}%
\rput(8,3.6){Re ($Z$)}%
\rput[l](4.4,6){\Large \darkgreen inductive}%
\rput[r](3.6,2){\Large \red capacitive}%
\rput(4.4,4.5){\Large $\frac{\lambda}{4}$}%
\rput(8,4.5){\Large $\frac{\lambda}{2}$}%
\end{pspicture}%
\caption{Impedance of a transmission line as a function of its length $l$}
\label{tangens}
\end{figure}
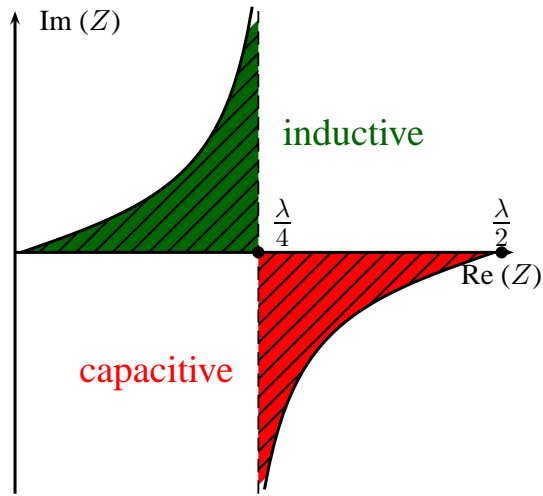

The peculiarity of a transmission line is that it behaves either as an inductance, a capacitor, or a resistor depending on its length. The impedance of such a line (if lossless!) is given by 
\begin{equation}
	Z_{\text{in}} = \text{j}Z_{\text{0}}\tan(\beta l)\hspace{0.2cm}.
\label{eq:11}
\end{equation}
The function in Eq.\,(\ref{eq:11}) has a pole at a transmission line length of $\lambda/4$ (Fig.\,\ref{tangens}). 
Therefore adding a transmission line with this length results in a change of $\Gamma$ by a factor $-$1:
\begin{equation}
	\Gamma_{\text{in}} = \Gamma_{\text{load}} \text{e}^{-\text{j}2\beta l} = \Gamma_{\text{load}} \text{e}^{-\text{j}2(\frac{2\pi}{\lambda}) l} \stackrel{l=\frac{\lambda}{4}}{=} \Gamma_{\text{load}} \text{e}^{-\text{j}\pi} = -\Gamma_{\text{load}}\hspace{0.2cm}.
\label{eq:12}
\end{equation}
Again this is equivalent to changing the original impedance $z$ to its admittance $1/z$ or the clockwise movement of the impedance vector by 180$^{\circ}$. Especially when starting with a short circuit (at $-$1 in the Smith chart), adding a transmission line of length $\lambda/4$ transforms it into an open circuit (at $+$1 in the Smith chart).

A line that is shorter than $\lambda/4$ behaves as an inductance, while a line that is longer acts as a capacitor.
Since these properties of transmission lines are used very often, the Smith chart usually has a ruler around its border, where one can read $l/\lambda$ --- it is the parametrization of the outermost circle.
\subsection{Examples of different 2-ports}
In general, the reflection coefficient when looking through a 2-port $\Gamma_{\text{in}}$ is given via the S-matrix of the 2-port and the reflection coefficient of the load $\Gamma_{\text{load}}$:
\begin{equation}
	\Gamma_{\text{in}} = \text{S}_{11} + \frac{\text{S}_{12} \text{S}_{21} \Gamma_{\text{load}}}{1 - \text{S}_{22} \Gamma_{\text{load}}}\hspace{0.2cm}.
\label{eq:13}
\end{equation}
In general, the outer circle of the Smith chart as well as its real axis are mapped to other circles and lines.

In the following three examples different 2-ports are given along with their S-matrix, and their representation in the Smith chart is discussed. For illustration, a simplified Smith chart consisting of the outermost circle and the real axis only is used for reasons of simplicity.
\subsubsection{Transmission line $\lambda/16$}
\label{tl}
\begin{figure}[htbp]
\centering\begin{pspicture}(0,0)(5,5)%
\pscircle[linecolor=red,linewidth=0.06](2.5,2.5){2.3}%
\psline[linestyle=dashed](0,2.5)(5,2.5)%
\psline[linewidth=0.06,linecolor=blue,ArrowInside=->,ArrowInsidePos=0.25,arrowscale=2](0.8,4)(4.2,1)%
\pscircle[linecolor=blue,fillstyle=solid](0.8,4){0.08}%
\pscircle[linecolor=blue,fillstyle=solid](2.5,2.5){0.08}%
\pscircle[linecolor=blue,fillstyle=solid](4.2,1){0.08}%
\psarc[arrows=<-,linewidth=0.06](2.5,2.5){2.5}{140}{180}%
\rput[l](1,4){\blue $z = 0$}%
\rput[l](2.7,2.6){\blue $z = 1$}%
\rput[l](4.4,1){\blue $z = \infty$}%
\pstextpath[c]{\psline[linestyle=none](0.6,3.8)(4,0.8)}{\blue increasing terminating}%
\pstextpath[c](0,-0.5){\psline[linestyle=none](0.6,3.8)(4,0.8)}{\blue resistor}%
\end{pspicture}%
\caption{Rotation of the reference plane of the Smith chart when adding a transmission line}
\label{tlsimple}
\end{figure}
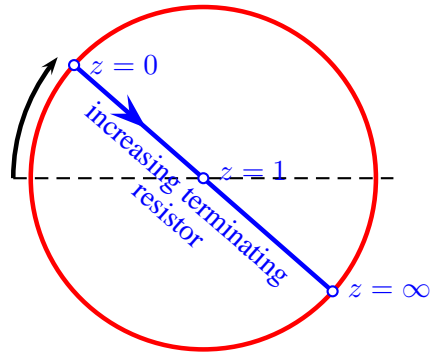

The S-matrix of a $\lambda/16$ transmission line is
\begin{equation}
	\text{S} = \left[
		\begin{array}{cc}
			0 & \text{e}^{-\text{j}\frac{\pi}{8}} \\
			\text{e}^{-\text{j}\frac{\pi}{8}} & 0 \\
		\end{array}
\right] 
\label{eq:14}
\end{equation}
with the resulting reflection coefficient 
\begin{equation}
	\Gamma_{\text{in}} = \Gamma_{\text{load}} \text{e}^{-\text{j}\frac{\pi}{4}}\hspace{0.2cm}.
\label{eq:15}
\end{equation}
This corresponds to a rotation of the real axis of the Smith chart by an angle of 45$^{\circ}$ (Fig.\,\ref{tlsimple}) and hence a change of the reference plane of the chart (Fig.\,\ref{tlsimple}). Consider, for example, a transmission line terminated by a short and hence $\Gamma_{\text{load}} = -1$.  The resulting reflection coefficient is then equal to $\Gamma_{\text{in}} = \text{e}^{-\text{j}\frac{\pi}{4}}$.
\subsubsection{Attenuator 3\,dB}
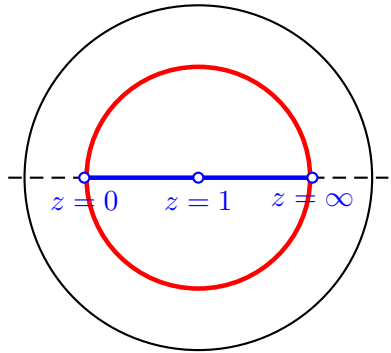
\begin{figure}[htbp]
\centering\begin{pspicture}(0,0)(5,5)%
\pscircle(2.5,2.5){2.3}%
\psline[linestyle=dashed](0,2.5)(5,2.5)%
\psline[linewidth=0.06,linecolor=blue](1,2.5)(4,2.5)%
\pscircle[linewidth=0.06,linecolor=red](2.5,2.5){1.5}%
\pscircle[linecolor=blue,fillstyle=solid](1,2.5){0.08}%
\pscircle[linecolor=blue,fillstyle=solid](2.5,2.5){0.08}%
\pscircle[linecolor=blue,fillstyle=solid](4,2.5){0.08}%
\rput(1,2.2){\blue $z = 0$}%
\rput(2.5,2.2){\blue $z = 1$}%
\rput(4,2.2){\blue $z = \infty$}%
\end{pspicture}%
\caption{Illustration of the appearance of an attenuator in the Smith chart}
\label{att}
\end{figure}

The S-matrix of an attenuator is given by
\begin{equation}
	\text{S} = \left[
		\begin{array}{cc}
			0 & \frac{\sqrt{2}}{2} \\
			\frac{\sqrt{2}}{2} & 0 \\
		\end{array}
\right]\hspace{0.2cm}. 
\label{eq:16}
\end{equation}
The resulting reflection coefficient is
\begin{equation}
	\Gamma_{\text{in}} = \frac{\Gamma_{\text{load}}}{2}\hspace{0.2cm}.
\label{eq:17}
\end{equation}
In the Smith chart, the connection of such an attenuator causes the outermost circle to shrink to a radius of 0.5\footnote{An attenuation of 3\,dB corresponds to a reduction by a factor 2 in power.} (Fig.\,\ref{att}).
\subsubsection{Variable load resistor}
\begin{figure}[htbp]
\centering\begin{pspicture}(0,0)(5,5)%
\psset{linewidth=0.06}
\pscircle[linecolor=red](2.5,2.5){2.3}%
\psline[linecolor=blue](0.2,2.5)(4.8,2.5)%
\pscircle[linecolor=blue,fillstyle=solid](0.2,2.5){0.08}%
\pscircle[linecolor=blue,fillstyle=solid](2.5,2.5){0.08}%
\pscircle[linecolor=blue,fillstyle=solid](4.8,2.5){0.08}%
\rput[r](0,2.5){\blue $z = 0$}%
\rput(2.5,2.2){\blue $z = 1$}%
\rput[l](5,2.5){\blue $z = \infty$}%
\end{pspicture}%
\caption{A variable load resistor in the simplified Smith chart. Since the impedance has a real part only, the signal remains on the real axis of the $\Gamma$ plane}
\label{res}
\end{figure}
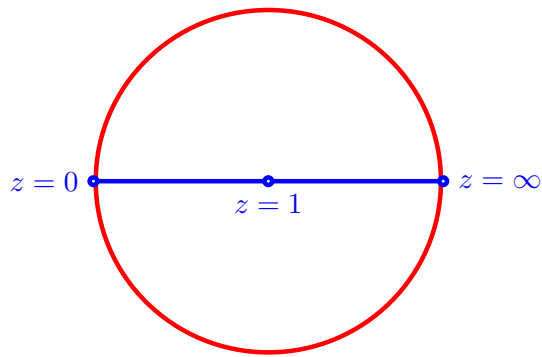

Adding a variable load resistor (0 $< z < \infty$) is the simplest case that can be depicted in the Smith chart. It means moving through the chart along its real axis (Fig.\,\ref{res}).
\section{Advantages of the Smith chart --- a summary}
\begin{itemize}
	\item The diagram offers a compact and handy representation of all passive impedances\footnote{Passive impedances are impedances with positive real part.} from 0 to $\infty$. Impedances with negative real part such as reflection amplifier or any other active device would show up outside the Smith chart. 

	\item Impedance mismatch is easily spotted in the chart.

	\item Since the mapping converts impedances or admittances ($y = \frac{1}{z}$) into reflection factors and vice versa, it is particularly interesting for studies in the radio frequency and microwave domain. For reasons of convenience, electrical quantities are usually expressed in terms of direct or forward waves and reflected or backwards waves in these frequency ranges instead of voltages and currents used at lower frequencies. 

	\item The transition between impedance and admittance in the chart is particularly easy: $\Gamma(\text{y = }\frac{1}{z}) = -\Gamma(z)$.

	\item Furthermore the reference plane in the Smith chart can be moved very easily by adding a transmission line of proper length (Section \ref{tl}).
	\item Many Smith charts have rulers below the complex $\Gamma$ plane from which a variety of quantities such as the return loss can be determined. For a more detailed discussion see Appendix \ref{appendix2}.
\end{itemize}
\section{Examples for applications of the Smith chart}
In this section two practical examples of common problems are given, where the use of the Smith chart greatly facilitates their solution.
\subsection{A step in characteristic impedance}
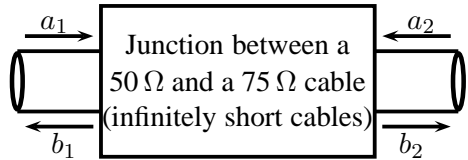
\begin{figure}[htbp]
\centering\begin{pspicture}(0,0)(6,4)%
\psset{linewidth=0.05}%
\pspolygon(1.2,1)(4.8,1)(4.8,3)(1.2,3)%
\psline(4.8,1.6)(5.9,1.6)%
\psline(4.8,2.4)(5.9,2.4)%
\psline(0.1,1.6)(1.2,1.6)%
\psline(0.1,2.4)(1.2,2.4)%
\psellipse[linewidth=0.06](5.9,2)(0.1,0.4)%
\psellipse[linewidth=0.06](0.1,2)(0.1,0.4)%
\rput(3,2.5){Junction between a}%
\rput(3,2){50\,$\Omega$ and a 75\,$\Omega$ cable}%
\rput(3,1.5){(infinitely short cables)}%
\pstextpath[c](0,0.1){\psline[arrows=->](0.2,2.6)(1.1,2.6)}{$a_{1}$}%
\pstextpath[c](0,-0.35){\psline[arrows=<-](0.2,1.4)(1.1,1.4)}{$b_{1}$}%
\pstextpath[c](0,0.1){\psline[arrows=<-](4.9,2.6)(5.8,2.6)}{$a_{2}$}%
\pstextpath[c](0,-0.35){\psline[arrows=->](4.9,1.4)(5.8,1.4)}{$b_{2}$}%
\end{pspicture}
\caption{Illustration of the junction between a coaxial cable with 50\,$\Omega$ characteristic impedance and another with 75\,$\Omega$ characteristic impedance respectively. Infinitely short cables are assumed -- only the junction is considered}
\label{junct}
\end{figure}

Consider a junction between two infinitely short cables, one with a characteristic impedance of 50\,$\Omega$ and the other with 75\,$\Omega$ (Fig.\,\ref{junct}). 
The waves running into each port are denoted with $a_{i}$ ($i = 1,2$) whereas the waves coming out of every point are denoted with $b_{i}$. The reflection coefficient for port 1 is then calculated as
\begin{equation}
	\Gamma_{1} = \frac{Z - Z_{1}}{Z + Z_{1}} = \frac{75 - 50}{75 + 50} = 0.2\hspace{0.2cm}.
\label{eq:18}
\end{equation}
Thus the voltage of the reflected wave at port 1 is 20\% of the incident wave ($a_{2} = a_{1}$ $\cdot$ $0.2$) and the reflected power at port 1 is 4\%\footnote{Power is proportional to $\Gamma ^{2}$ and thus 0.2$^{2}$ = 0.04.}. From conservation of energy, the transmitted power has to be 96\% and it follows that $b_{2}^{2}$ = 0.96.

A peculiarity here is that the transmitted energy is \textit{higher} than the energy of the incident wave, since $E_{\text{incident}}$ = 1, $E_{\text{reflected}}$ = 0.2 and therefore $E_{\text{transmitted}}$ = $E_{\text{incident}}$ + $E_{\text{reflected}}$ = 1.2. The transmission coefficient $t$ is thus $t$ = 1 + $\Gamma$. Also note that this structure is not symmetric (S$_{11} \neq $ S$_{22}$), but only reciprocal (S$_{21} = $ S$_{12}$).

\begin{figure}[htbp]
\centering\input{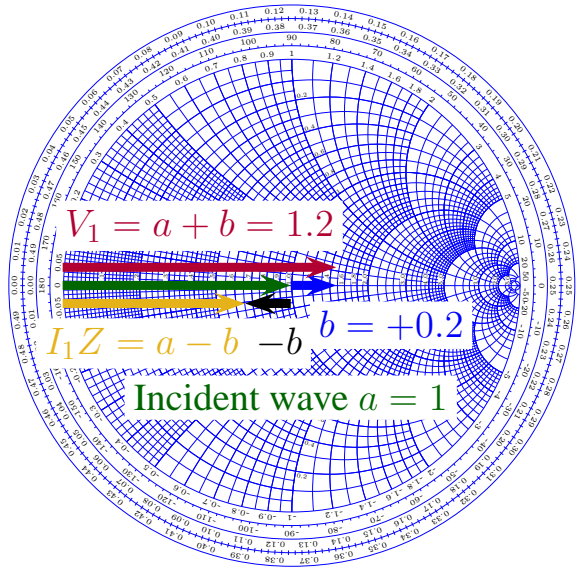}
\caption{Visualization of the two-port formed by the two cables of different characteristic impedance}
\label{tlsolution}
\end{figure}

The visualization of this structure in the Smith chart is easy, since all impedances are real and thus all vectors are located on the real axis (Fig.\,\ref{tlsolution}).

As stated before, the reflection coefficient is defined with respect to voltages. For currents its sign inverts and thus a positive reflection coefficient in terms of voltage definition becomes negative when defined with respect to current.

\begin{figure}[htbp]
\centering\input{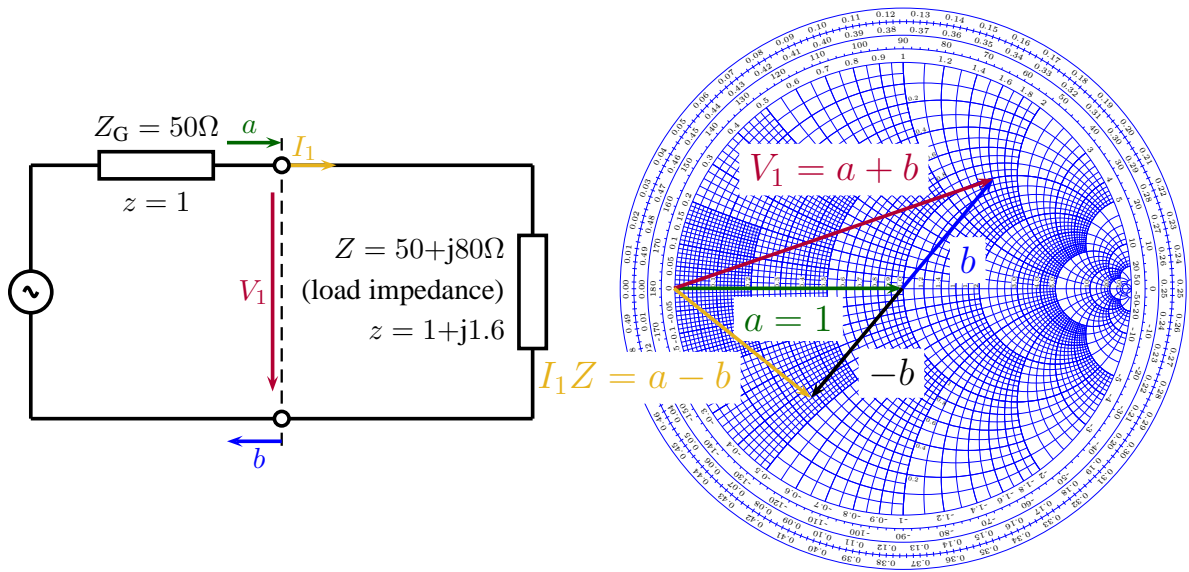}
\caption{Visualization of the two-port depicted on the left in the Smith chart}
\label{generaltl}
\end{figure}

For a more general case, e.g., $Z_{1}$ = 50\,$\Omega$ and $Z_{2}$ = 50 + $\text{j}$80\,$\Omega$, the vectors in the chart are depicted in Fig.\,\ref{generaltl}.
\subsection{Determination of the $Q$ factors of a cavity}
One of the most common cases where the Smith chart is used is the determination of the quality factor of a cavity. This section is dedicated to the illustration of this task.

\begin{figure}[htbp]
\centering\begin{pspicture}(0,0)(12,5.3)%
%
\rput(1.3,0){%
%
\psline(0,1)(1,1)%
\psline(1,1)(1,1.6)%
\psline(0,1)(0,1.4)%
\pscircle(0,1.8){0.4}%
\rput(-0.15,1.8){%
\psplot[algebraic=true]{0}{0.3}{0.1*sin(20.944*x)}%
}%
\psline(0,2.2)(0,2.8)%
\pspolygon(-0.2,2.8)(0.2,2.8)(0.2,3.8)(-0.2,3.8)%
\psline(0,3.8)(0,4.2)%
\psline(0,4.2)(1,4.2)%
\psline(1,4.2)(1,3.6)%
\pszigzag[coilwidth=0.3,coilheight=0.5,coilarm=0.01](1,1.6)(1,3.6)%
\rput[r](-0.4,3.3){$Z_{\text{G}}$}%
%
\psline(1.6,1)(1.6,1.6)%
\psline(1.6,1)(10,1)%
\psline(1.6,4.2)(1.6,3.6)%
\psline(1.6,4.2)(10,4.2)%
\pszigzag[coilwidth=0.3,coilheight=0.5,coilarm=0.01](1.6,1.6)(1.6,3.6)%
%
\psline(4,1)(4,1.8)%
\psline(4,4.2)(4,3.4)%
\pspolygon(3.8,1.8)(4.2,1.8)(4.2,3.4)(3.8,3.4)%
\rput(4.4,2.6){$R$}%
%
\psline(5.3,1)(5.3,1.8)%
\psline(5.3,4.2)(5.3,3.4)%
\pscoil[coilwidth=0.3,coilheight=0.5,coilarm=0.01](5.3,1.8)(5.3,3.4)%
\rput[l](5.6,2.6){$L$}%
%
\psline(6.6,1)(6.6,2.5)%
\psline(6.6,4.2)(6.6,2.7)%
\psline(6.3,2.5)(6.9,2.5)%
\psline(6.3,2.7)(6.9,2.7)%
\rput[l](7.1,2.6){$C$}%
%
\psline[linestyle=dashed,linecolor=green](0.5,0.4)(0.5,4.8)%
\pscircle[fillstyle=solid](0.5,4.2){0.1}%
\pscircle[fillstyle=solid](0.5,1){0.1}
\psline[linestyle=dashed,linecolor=green](9,0.4)(9,4.8)%
\pscircle[fillstyle=solid](9,1){0.1}%
\pscircle[fillstyle=solid](9,4.2){0.1}%
}%
%
\rput(-0.7,0){%
\psline[arrows=->](11.6,4)(11.6,1.2)%
\rput[l](11.8,2.6){$V_{\text{beam}}$}%
\psline[arrows=->](1.4,2.4)(1.4,1.2)%
\rput[r](1.2,1.9){$V_{0}$}%
}%
\psline[linewidth=0.1,arrows=->,arrowscale=1](1,4.7)(1.7,4.7)%
\rput[r](0.9,4.7){$Z_{\text{input}}$}%
\psline[linewidth=0.1,arrows=<-,arrowscale=1](10.4,4.7)(11.1,4.7)%
\rput[l](11.2,4.7){$Z_{\text{shunt}}$}%
\end{pspicture}
\caption{The equivalent circuit that can be used to describe a cavity. The transformer is hidden in the coupling of the cavity ($Z \approx 1$\,M$\Omega$, seen by the beam) to the generator (usually $Z = 50\,\Omega$)}
\label{rlc}
\end{figure}
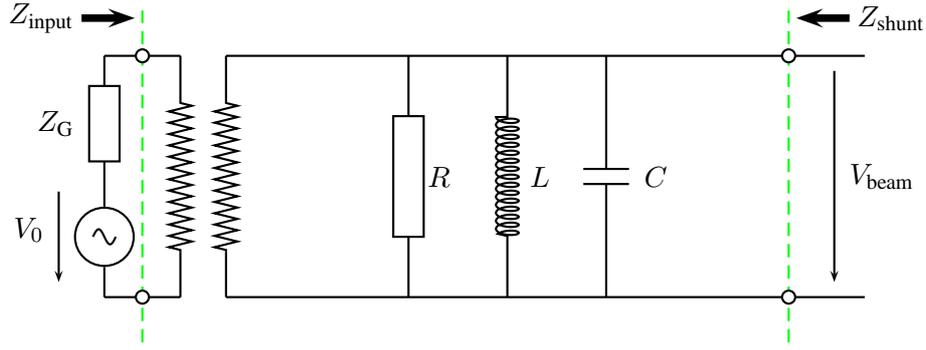

A cavity can be described by a parallel $RLC$ circuit (Fig.\,\ref{rlc})
where the resonance condition is given when:
\begin{equation}
	\omega L = \frac{1}{\omega C}\hspace{0.2cm}.
\label{eq:19}
\end{equation}
This leads to the resonance frequency of
\begin{equation}
	\omega_{\text{res}} = \frac{1}{\sqrt{LC}} \text{\hspace{1cm} or \hspace{1cm} }f_{\text{res}} = \frac{1}{2 \pi}\frac{1}{\sqrt{LC}}\hspace{0.2cm}. 
\label{eq:20}
\end{equation}

The Impedance $Z$ of such an equivalent circuit is given by
\begin{equation}
	Z(\omega) = \frac{1}{\frac{1}{R} + \text{j}\omega C + \frac{1}{\text{j}\omega L}}\hspace{0.2cm}.
\label{eq:24}
\end{equation}

\begin{figure}[htbp]
\centering\begin{pspicture}(0,0)(10,7)%
\newrgbcolor{darkyellow}{0.9 0.7 0.1}%
\psset{linewidth=0.06}
\rput(1,0){%
%
%
\psline[arrows=->](0.2,3.5)(8.8,3.5)%
%
%
\psline[arrows=->](0.8,0.2)(0.8,6.8)%
%
%
\psline[linewidth=0.03,linestyle=dashed,linecolor=blue](3.3,1)(3.3,6)%
\psline[linewidth=0.03,linecolor=blue](0.8,3.5)(3.3,1)%
%
%
\psline[arrows=->,linecolor=blue](0.8,3.5)(3.3,6)%
%
%
\pscircle[linecolor=green
](3.3,3.5){2.5}%
\psarc[arrows=<-,linecolor=green,linewidth=0.04,arrowscale=2.5](3.3,3.5){2.5}{135}{180}%
\psarc[arrows=<-,linecolor=green,linewidth=0.04,arrowscale=2.5](3.3,3.5){2.5}{45}{90}%
\psarc[arrows=<-,linecolor=green,linewidth=0.04,arrowscale=2.5](3.3,3.5){2.5}{315}{0}%
\psarc[arrows=<-,linecolor=green,linewidth=0.04,arrowscale=2.5](3.3,3.5){2.5}{225}{270}%
\pscircle[linecolor=green,fillstyle=solid](0.8,3.5){0.1}%
\pscircle[linecolor=green,fillstyle=solid](5.8,3.5){0.1}%
\pscircle[linecolor=green,fillstyle=solid](3.3,6){0.1}%
\pscircle[linecolor=green,fillstyle=solid](3.3,1){0.1}%
%
%
\psarc[linecolor=blue,arrows=<->,linewidth=0.045](0.8,3.5){1.5}{0}{45}%
%
%
\rput(8.8,3.1){Re ($Z$)}%
\rput(1.5,6.6){Im ($Z$)}%
\rput[r](2,4){\blue $45^{\circ}$}%
}%
%
%
\psline[arrows=->,linecolor=darkyellow,linewidth=0.045](5.35,6.2)(4.3,6)%
\psline[arrows=->,linecolor=darkyellow,linewidth=0.045](7.4,4)(6.8,3.5)%
\psline[arrows=->,linecolor=darkyellow,linewidth=0.045](5.35,0.8)(4.3,1)%
\psline[arrows=->,linecolor=darkyellow,linewidth=0.045](1.4,2.4)(1.8,3.5)%
\psline[arrows=->,linecolor=darkyellow,linewidth=0.045](1.4,4.6)(1.8,3.5)%
\rput[l](4.4,6.6){\psframebox[fillstyle=solid,linewidth=0.03]{$f = f^{\,-}_{(-3\text{dB})}$}}%
\rput[l](7.4,4){\psframebox[fillstyle=solid,linewidth=0.03]{$f = f_{(\text{res})}$}}%
\rput[l](4.4,0.4){\psframebox[fillstyle=solid,linewidth=0.03]{$f = f^{\,+}_{(-3\text{dB})}$}}%
\rput[r](1.4,4.6){\psframebox[fillstyle=solid,linewidth=0.03]{$f = 0$}}%
\rput[r](1.4,2.4){\psframebox[fillstyle=solid,linewidth=0.03]{$f \rightarrow \infty$}}%
\end{pspicture}
\caption{Schematic drawing of the 3\,dB bandwidth in the impedance plane}
\label{3db}
\end{figure}
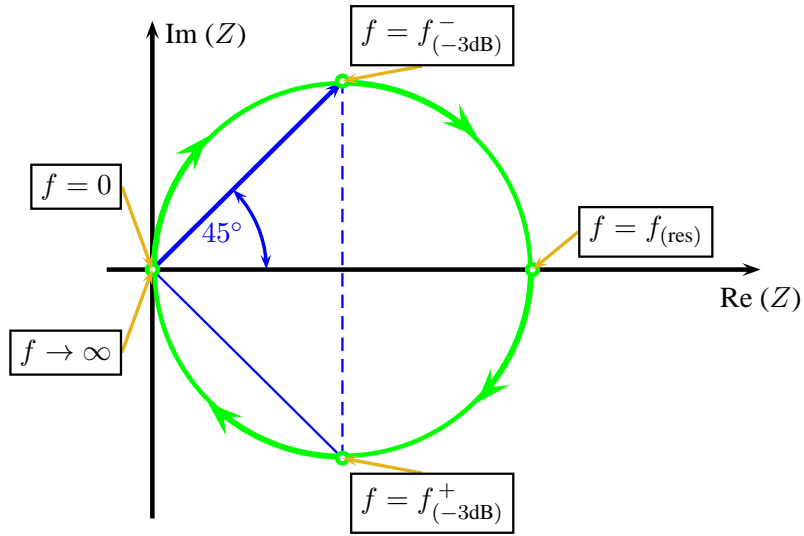

The 3\,dB bandwidth $\Delta f$ refers to the points where Re($Z$) = Im($Z$) which corresponds to two vectors with an argument of 45$^{\circ}$ (Fig.\,\ref{3db}) and an impedance of $|Z_{(-3\text{dB})}| = 0.707 R = R/\sqrt{2}$.

In general, the quality factor $Q$ of a resonant circuit is defined as the ratio of the stored energy $W$ over the energy dissipated in one cycle $P$:
\begin{equation}
	Q = \frac{\omega W}{P}\hspace{0.2cm}.
\label{eq:21}
\end{equation}
The $Q$ factor for a resonance can be calculated via the 3\,dB bandwidth and the resonance frequency:
\begin{equation}
	Q = \frac{f_{\text{res}}}{\Delta f}\hspace{0.2cm}.
\label{eq:22}
\end{equation}
For a cavity, three different quality factors are defined:
\begin{itemize}
	\item $Q_{0}$ (unloaded $Q$): $Q$ factor of the unperturbed system, i.\,e., the stand alone cavity;
	\item $Q_{\text{L}}$ (loaded $Q$): $Q$ factor of the cavity when connected to generator and measurement circuits;
	\item $Q_{\text{ext}}$ (external $Q$): $Q$ factor that describes the degeneration of $Q_{0}$ due to the generator and diagnostic impedances.
\end{itemize}
All these $Q$ factors are connected via a simple relation:
\begin{equation}
	\frac{1}{Q_{\text{L}}} = \frac{1}{Q_{0}} + \frac{1}{Q_{\text{ext}}}\hspace{0.2cm}.
\label{eq:23}
\end{equation}
The coupling coefficient $\beta$ is then defined as
\begin{equation}
	\beta = \frac{Q_{0}}{Q_{\text{ext}}}\hspace{0.2cm}.
\label{eq:25}
\end{equation}
This coupling coefficient is not to be confused with the propagation coefficient of transmission lines which is also denoted as $\beta$.

In the Smith chart, a resonant circuit shows up as a circle (Fig.\,\ref{qfactor}, circle shown in the detuned short position). The larger the circle, the stronger the coupling. Three types of coupling are defined depending on the range of $beta$ (= size of the circle, assuming the circle is in the detuned short position):
\begin{figure}[htbp]
\centering\input{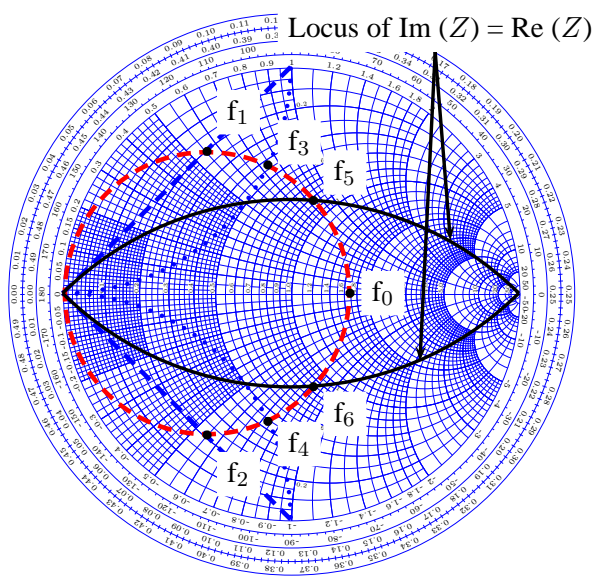}
\caption{Illustration of how to determine the different $Q$ factors of a cavity in the Smith chart}
\label{qfactor}
\end{figure}
\begin{itemize}
	\item Undercritical coupling ($0 < \beta < 1$): The radius of resonance circle is smaller than 0.25. Hence the centre of the chart lies outside the circle. 
	\item Critical coupling ($\beta = 1$): The radius of the resonance circle is exactly 0.25. Hence the circle touches the centre of the chart. 
	\item Overcritical coupling ($1 < \beta < \infty$): The radius of the resonance circle is larger than 0.25. Hence the centre of the chart lies inside the circle.
\end{itemize}
In practice, the circle may be rotated around the origin due to the transmission lines between the resonant circuit and the measurement device.

From the different marked frequency points in Fig.\,\ref{qfactor} the 3\,dB bandwidth and thus the quality factors $Q_{0}$, $Q_{\text{L}}$ and $Q_{\text{ext}}$ can be determined as follows:
\begin{itemize}
\item The unloaded $Q$ can be determined from f$_{5}$ and f$_{6}$. The condition to find these points is Re($Z$) = Im($Z$) with the resonance circle in the detuned short position.
\item The loaded $Q$ can be determined from f$_{1}$ and f$_{2}$. The condition to find these points is $\left|\text{Im}(\text{S}_{11})\right| \rightarrow$ max.
\item The external $Q$ can be calculated from f$_{3}$ and f$_{4}$. The condition to determine these points is $Z$ = $\pm \text{j}$.
\end{itemize}

To determine the points f$_{1}$ to f$_{6}$ with a network analyzer, the following steps are applicable:
\begin{itemize}
	\item f$_{1}$ and f$_{2}$: Set the marker format to Re(S$_{11}$) + j\,Im(S$_{11}$) and determine the two points, where Im(S$_{11}$) = max.
	\item f$_{3}$ and f$_{4}$: Set the marker format to $Z$ and find the two points where $Z = \pm$j.
	\item f$_{5}$ and f$_{6}$: Set the marker format to $Z$ and locate the two points where Re($Z$) = Im($Z$).
\end{itemize}
%
%
\appendix
\section{Transformation of lines with constant real or imaginary part from the impedance plane to the $\Gamma$ plane}
\label{appendix1}
This section is dedicated to a detailed calculation of the transformation of coordinate lines form the impedance to the $\Gamma$ plane. The interested reader is referred to Ref.\,\cite{paul1969} for a more detailed study.

\begin{figure}[htbp]
\centering\begin{pspicture}(0,0)(6,6)%
\rput(0,0.5){%
\psaxes[showorigin=false,tickstyle=bottom,labelsep=5pt]{->}(5.5,5.5)%
\psline[linestyle=dotted,linewidth=0.01](0,1)(5,1)%
\psline[linestyle=dotted,linewidth=0.01](0,2)(5,2)%
\psline[linestyle=dotted,linewidth=0.01](0,3)(5,3)%
\psline[linestyle=dotted,linewidth=0.01](0,4)(5,4)%
\psline[linestyle=dotted,linewidth=0.01](0,5)(5,5)%
\psline[linestyle=dotted,linewidth=0.01](1,0)(1,5)%
\psline[linestyle=dotted,linewidth=0.01](2,0)(2,5)%
\psline[linestyle=dotted,linewidth=0.01](3,0)(3,5)%
\psline[linestyle=dotted,linewidth=0.01](4,0)(4,5)%
\psline[linestyle=dotted,linewidth=0.01](5,0)(5,5)%
\rput[l](5.8,0){Re($z$)}
\rput[l](0.3,5.3){Im($z$)}
\psline[linecolor=blue,arrows=->](0,0)(3.5,3)%
\pscircle[linecolor=blue,fillcolor=blue,fillstyle=solid](3.5,3){0.05}%
\rput(3.5,3.2){\large \blue $z = 3.5 + $j$3$}
}%
\end{pspicture}%
\caption{The complex impedance plane}
\label{axis}
\end{figure}
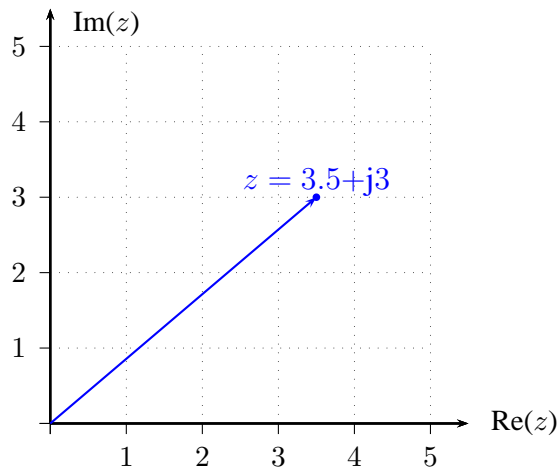

Consider a coordinate system in the complex impedance plane. The real part $R$ of each impedance is assigned to the horizontal axis and the imaginary part $X$ of each impedance to the vertical axis (Fig.\,\ref{axis}).
For reasons of simplicity, all impedances used in this calculation are normalized to an impedance $Z_{\text{0}}$. This leads to the simplified transformation between impedance and $\Gamma$ plane:
\begin{equation}
	\Gamma = \frac{z - 1}{z + 1}\hspace{0.2cm}. 
\label{eq:a1}
\end{equation}
$\Gamma$ is a complex number itself: $\Gamma =  a + $j$c$. Using this identity and substituting $z= R +$ j$X$ in equation (\ref{eq:a1}) one obtains
\begin{equation}
	\Gamma = \frac{z - 1}{z + 1} = \frac{R+\text{j}X - 1}{R+\text{j}X + 1} = a + \text{j}c\hspace{0.2cm}.
\label{eq:a2}
\end{equation}
From this the real and the imaginary part of $\Gamma$ can be calculated in terms of $a$, $c$, $R$ and $X$:
\begin{eqnarray}
	\text{Re: }a(R + 1) - cX &=& R - 1 ;\\
	\text{Im: }c(R + 1) + aX &=& X .
\end{eqnarray}
\subsection{Lines with constant real part}
To consider lines with constant real part, one can extract an expression for $X$ from Eq.\,(A.4):
\begin{equation}
	X = c\frac{1 + R}{1 - a}
\label{eq:a5}
\end{equation}
and substitute this into Eq.\,(A.3):
\begin{equation}
	a^{2} + c^{2} - 2a\frac{R}{1 + R} + \frac{R - 1}{R + 1} = 0\hspace{0.2cm}.
\label{eq:a6}
\end{equation}
Completing the square, one obtains the equation of a circle:
\begin{equation}
	\left(a - \frac{R}{1 + R} \right)^2 + c^2 = \frac{1}{(1 + R)^2}\hspace{0.2cm}.
\label{eq:a7}
\end{equation}
From this equation the following properties can be deduced:
\begin{itemize}
	\item The centre of each circle lies on the real $a$ axis.
	\item Since $\frac{R}{1 + R} \geq 0$, the centre of each circle lies on the positive real $a$ axis.
	\item The radius $\rho$ of each circle follows the equation $\rho = \frac{1}{(1 + R)^2} \leq 1$.
	\item The maximal radius is 1 for $R$ = 0.
\end{itemize}
\subsubsection{Examples}
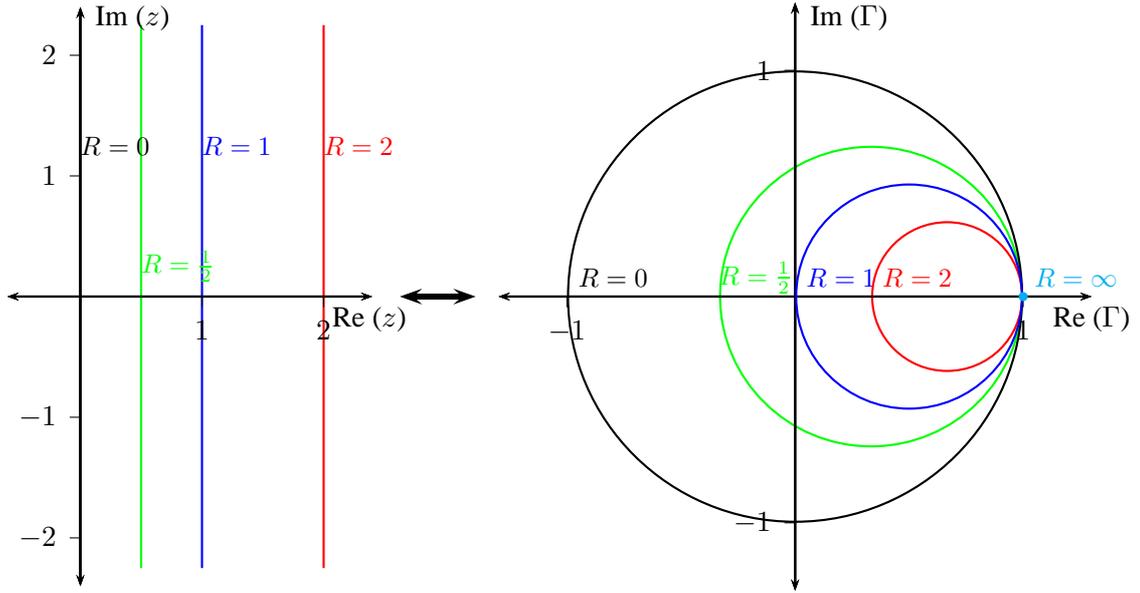
\begin{figure}[htbp]
\centering\begin{pspicture}(0,0)(15,8)%
\rput(6.6,0){%
\rput[l](1.15,4.25){\small $R = 0$}%
\rput[l](3,4.25){\small \green $R = \frac{1}{2}$}%
\rput[l](4.15,4.25){\small \blue $R = 1$}%
\rput[l](5.15,4.25){\small \red $R = 2$}%
\rput[l](7.15,4.25){\small \cyan $R = \infty$}%
\rput[l](4.2,7.7){Im ($\Gamma$)}%
\rput(7.9,3.7){Re ($\Gamma$)}
\rput(4,4){%
\pscircle[linecolor=green](1,0){2}%
\pscircle[linecolor=red](2,0){1}%
\psset{unit=3cm}%
\psaxes[showorigin=false,tickstyle=bottom,labelsep=5pt]{<->}(0,0)(-1.3,-1.3)(1.3,1.3)%
\pscircle(0,0){1}%
\pscircle[linecolor=blue](0.5,0){0.5}%
\pscircle[linecolor=cyan,fillcolor=cyan,fillstyle=solid](1,0){0.02}%
}%
}%
\psline[linecolor=green](2,0.4)(2,7.6)%
\psline[linecolor=blue](2.8,0.4)(2.8,7.6)%
\psline[linecolor=red](4.4,0.4)(4.4,7.6)%
\rput[l](1.2,6){\small $R = 0$}%
\rput[l](2,4.4){\small \green $R = \frac{1}{2}$}%
\rput[l](2.8,6){\small \blue $R = 1$}%
\rput[l](4.4,6){\small \red $R = 2$}%
\rput(1.2,4){%
\psset{unit=1.6cm}%
\psaxes[showorigin=false,tickstyle=bottom,labelsep=5pt]{<->}(0,0)(-0.6,-2.4)(2.4,2.4)%
}%
\psline[arrows=<->,linewidth=0.08](5.4,4)(6.4,4)%
\rput[l](1.4,7.7){Im ($z$)}%
\rput(5,3.7){Re ($z$)}
\end{pspicture}%
\caption{Lines of constant real part transformed into the $\Gamma$ plane}
\label{rconst}
\end{figure}

Here the circles for different $R$ values are calculated and depicted graphically to illustrate the transformation from the $z$ to the $\Gamma$ plane.
\begin{enumerate}
	\item $R$ = 0: This leads to the centre coordinates (c$_{a}$/c$_{c}$) = $\left( \frac{0}{1 + 0}/0\right) = (0 / 0)$, $\rho = \frac{1}{1 + 0} = 1$
	\item $R$ = 0.5: (c$_{a}$/c$_{c}$) = $\left( \frac{0.5}{1 + 0.5}/0\right) = (\frac{1}{3} / 0)$, $\rho = \frac{1}{1 + 0.5} = \frac{2}{3}$
	\item $R$ = 1: (c$_{a}$/c$_{c}$) = $\left( \frac{1}{1 + 1}/0\right) = (\frac{1}{2} / 0)$, $\rho = \frac{1}{1 + 1} = \frac{1}{2}$
	\item $R$ = 2: (c$_{a}$/c$_{c}$) = $\left( \frac{2}{1 + 2}/0\right) = (\frac{2}{3} / 0)$, $\rho = \frac{1}{1 + 2} = \frac{1}{3}$
	\item $R$ = $\infty$: (c$_{a}$/c$_{c}$) = $\left( \frac{\infty}{1 + \infty}/0\right) = (1 / 0)$, $\rho = \frac{1}{1 + \infty} = 0$
\end{enumerate}
This leads to the circles depicted in Fig.\,\ref{rconst}.
\subsection{Lines with constant imaginary part}
To calculate the circles in the Smith chart that correspond to the lines of constant imaginary part in the impedance plane, the formulas (A.3) and (A.4) are used again. Only this time an expression for $R$ and $R$ + 1 is calculated from Eq.\,(A.3) 
\begin{equation}
	R = \frac{a + 1 - cX}{1 - a}\hspace{0.5cm}\text{ and }\hspace{0.5cm} 1 + R = \frac{2 - cX}{1 - a}
\label{eq:a8}
\end{equation}
and substituted into Eq.\,(A.4):
\begin{equation}
	a^{2} - 2a + 1 + c^{2} - 2\frac{c}{X} = 0\hspace{0.2cm}.
\label{eq:a9}
\end{equation}
Completing the square again leads to the equation of a circle:
\begin{equation}
	(a - 1)^{2} + \left(c - \frac{1}{X} \right)^{2} = \frac{1}{X^{2}}\hspace{0.2cm}.
\label{eq:a10}
\end{equation}
Examining this equation, the following properties can be deduced:
\begin{itemize}
	\item The centre of each circle lies on an axis parallel to the imaginary axis at a distance of 1.
	\item The first coordinate of each circle centre is 1.
	\item The second coordinate of each circle centre is $\frac{1}{X}$. It can be smaller or bigger than 0 depending on the value of $X$.
	\item No circle intersects the real a axis.
	\item The radius $\rho$ of each circle is $\rho = \frac{1}{\left|X\right|}$.
	\item All circles contain the point (1/0).
\end{itemize}
\subsubsection{Examples}
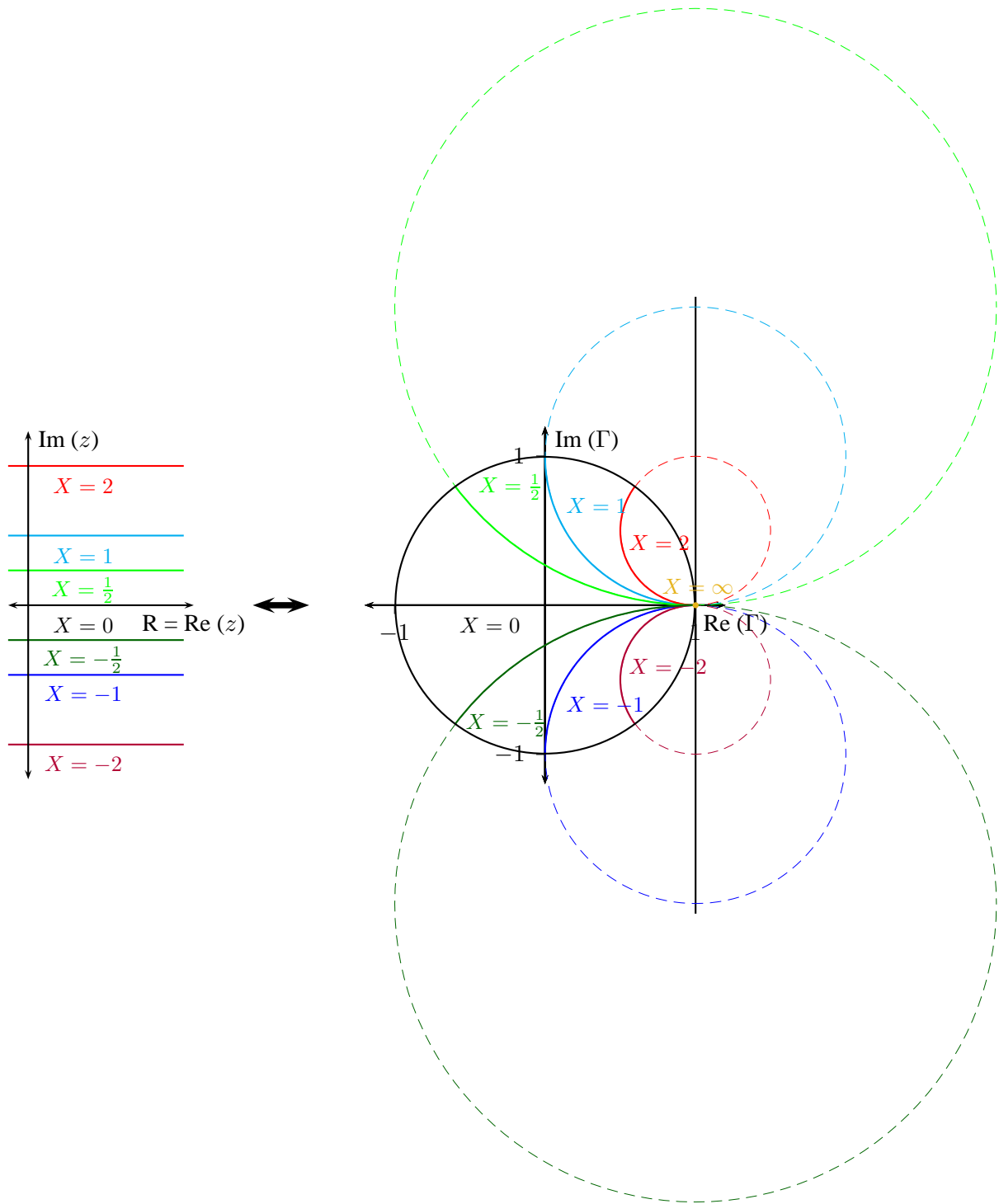
\begin{figure}[htbp]
\centering\begin{pspicture}(0,0)(15.8,19.2)%
\newrgbcolor{darkgreen}{0.0 0.4 0.0}%
\newrgbcolor{darkred}{0.7 0.0 0.2}%
\newrgbcolor{darkyellow}{0.9 0.7 0.1}%
\rput(8.56,9.6){%
\psset{unit=2.4cm}%
\psaxes[showorigin=false,tickstyle=bottom,labelsep=5pt]{<->}(0,0)(-1.2,-1.2)(1.2,1.2)%
}%
\psset{unit=0.8cm}
\rput(0,6){%
%
%
\rput(-1.3,0){%
%
%
\psline(15,-0.2)(15,12.2)%
\pscircle[linestyle=dashed,linecolor=red,linewidth=0.01](15,7.5){1.5}%
\pscircle[linestyle=dashed,linecolor=cyan,linewidth=0.01](15,9){3}%
\pscircle[linestyle=dashed,linecolor=green,linewidth=0.01](15,12){6}%
\pscircle[linestyle=dashed,linecolor=blue,linewidth=0.01](15,3){3}%
\pscircle[linestyle=dashed,linecolor=darkred,linewidth=0.01](15,4.5){1.5}%
\pscircle[linestyle=dashed,linecolor=darkgreen,linewidth=0.01](15,0){6}%
\psarcAB[linecolor=cyan](15,9)(12,9)(15,6)%
\psarcAB[linecolor=red](15,7.5)(13.8,8.4)(15,6)%
\psarcAB[linecolor=green](15,12)(10.2,8.4)(15,6)%
\psarcAB[linecolor=blue](15,3)(15,6)(12,3)%
\psarcAB[linecolor=darkred](15,4.5)(15,6)(13.8,3.6)%
\psarcAB[linecolor=darkgreen](15,0)(15,6)(10.2,3.6)%
\pscircle(12,6){3}%
\pscircle[linecolor=darkyellow,fillcolor=darkyellow,fillstyle=solid](15,6){0.06}%
\rput[l](12.2,9.3){Im ($\Gamma$)}
\rput(15.8,5.6){Re ($\Gamma$)}
}%
}%
%
%
\psset{unit=1cm}%
\rput[l](9.9,10.6){\small \red $X = 2$}%
\rput[l](8.9,11.2){\small \cyan $X = 1$}%
\rput[l](7.5,11.5){\small \green $X = \frac{1}{2}$}%
\rput[l](7.3,7.7){\small \darkgreen $X = -\frac{1}{2}$}%
\rput[l](8.9,8){\small \blue $X = -1$}%
\rput[l](9.9,8.6){\small \darkred $X = -2$}%
\rput[c](11,9.9){\small \darkyellow $X = \infty$}

\psset{unit=0.8cm}%
\rput(0,6){%
\rput[l](9,5.6){\small $X = 0$}%
%
%
\psline[linecolor=green](0,6.7)(3.5,6.7)%
\psline[linecolor=cyan](0,7.4)(3.5,7.4)%
\psline[linecolor=red](0,8.8)(3.5,8.8)%
\psline[linecolor=darkgreen](0,5.3)(3.5,5.3)%
\psline[linecolor=blue](0,4.6)(3.5,4.6)%
\psline[linecolor=darkred](0,3.2)(3.5,3.2)%
\psline[arrows=->](0.4,6)(0.4,9.5)%
\psline[arrows=->](0.4,6)(0.4,2.5)%
\psline[arrows=<->](0,6)(3.7,6)%
\rput[l](0.6,9.3){Im ($z$)}%
\rput(3.7,5.6){R = Re ($z$)}%
\rput(-2,2.4){%
%
%
\rput(3.5,6){\small \red $X = 2$}%
\rput(3.5,4.6){\small \cyan $X = 1$}%
\rput(3.5,3.9){\small \green $X = \frac{1}{2}$}%
\rput(3.5,3.2){\small $X = 0$}
\rput(3.5,2.5){\small \darkgreen $X = -\frac{1}{2}$}%
\rput(3.5,1.8){\small \blue $X = -1$}%
\rput(3.5,0.4){\small \darkred $X = -2$}%
}%
}%
\psset{unit=1cm}
%
%
\psline[arrows=<->,linewidth=0.1](3.9,9.6)(4.8,9.6)%
\end{pspicture}%
\caption{Lines of constant imaginary part transformed into the $\Gamma$ plane}
\label{xconst}
\end{figure}

In the following, examples for different $X$ values are calculated and depicted graphically to illustrate the transformation of the lines with constant imaginary part in the impedance plane to the corresponding circles in the $\Gamma$ plane.
\begin{enumerate}
	\item $X$ = -2: (c$_{a}$/c$_{c}$) = $\left( 1 / \frac{1}{-2}\right) = (1 / -0.5)$, $\rho = \frac{1}{\left|-2\right|} = 0.5$
	\item $X$ = -1: (c$_{a}$/c$_{c}$) = $\left( 1 / \frac{1}{-1}\right) = (1 / -1)$, $\rho = \frac{1}{\left|-1\right|} = 1$
	\item $X$ = -0.5: (c$_{a}$/c$_{c}$) = $\left( 1 / \frac{1}{-0.5}\right) = (1 / -2)$, $\rho = \frac{1}{\left|-2\right|} =2$
	\item $X$ = 0: (c$_{a}$/c$_{c}$) = $\left( 1 / \frac{1}{0}\right) = (1 / \infty)$, $\rho = \frac{1}{\left|0\right|} = \infty$ = real $a$ axis
	\item $X$ = 0.5: (c$_{a}$/c$_{c}$) = $\left( 1 / \frac{1}{0.5}\right) = (1 / 2)$, $\rho = \frac{1}{\left|-2\right|} = 2$
	\item $X$ = 1: (c$_{a}$/c$_{c}$) = $\left( 1 / \frac{1}{1}\right) = (1 / 1)$, $\rho = \frac{1}{\left|1\right|} = 1$
	\item $X$ = 2: (c$_{a}$/c$_{c}$) = $\left( 1 / \frac{1}{2}\right) = (1 / 0.5)$, $\rho = \frac{1}{\left|2\right|} = 0.5$
	\item $X$ = $\infty$: (c$_{a}$/c$_{c}$) = $\left( 1 / \frac{1}{\infty}\right) = (1 / 0)$, $\rho = \frac{1}{\left|\infty\right|} = 0$
\end{enumerate}
A graphical representation of the circles corresponding to these values is given in Fig.\,\ref{xconst}.
\newpage
\section{Rulers around the Smith chart}
\label{appendix2}
Some Smith charts provide rulers at the bottom to determine other quantities besides the reflection coefficient such as the return loss, the attenuation, the reflection loss etc. A short instruction on how to use these rulers as well as a specific example for such a set of rulers is given here.
\subsection{How to use the rulers}
First, one has to take the modulus (= distance between the centre of the Smith chart and the point in the chart referring to the impedance in question) of the reflection coefficient of an impedance either with a conventional ruler or, better, using a compass. Then refer to the coordinate denoted as CENTRE and go to the left or for the other part of the rulers to the right (except for the lowest line which is marked ORIGIN at the left which is the reference point of this ruler). The value in question can then be read from the corresponding scale.
\subsection{Example of a set of rulers}
\begin{figure}[htbp]
\centering\includegraphics[width=.9\linewidth]{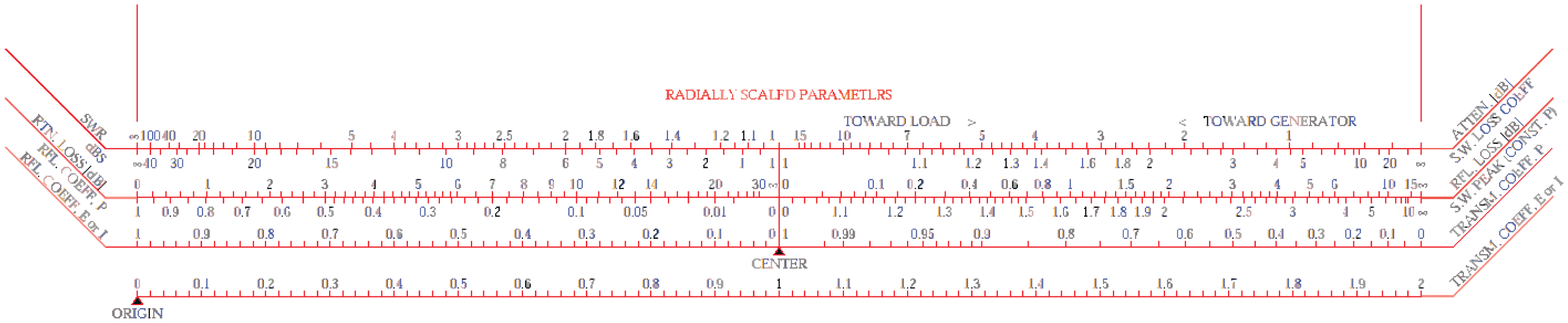}
\caption{Example for a set of rulers that can be found underneath the Smith chart}
\label{ruler}
\end{figure}

\begin{figure}[htbp]
\centering\includegraphics[width=.9\linewidth]{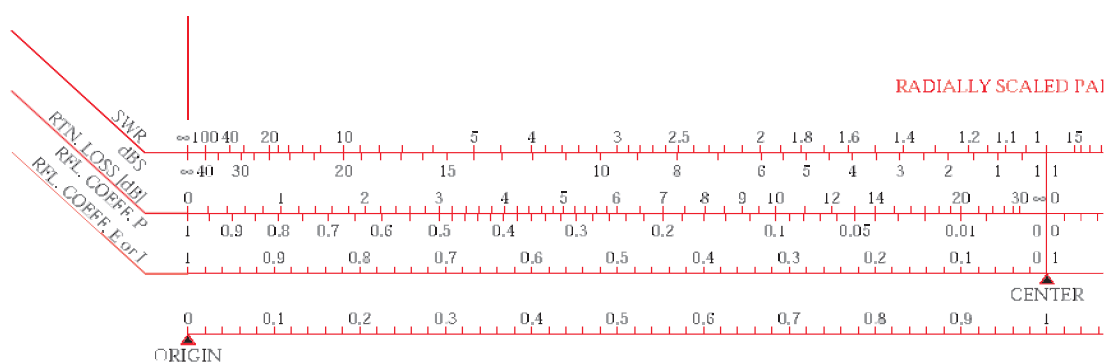}
\caption{Left part of the rulers depicted in Fig.\,\ref{ruler}}
\label{rulerleft}
\end{figure}

\begin{figure}[htbp]
\centering\includegraphics[width=.9\linewidth]{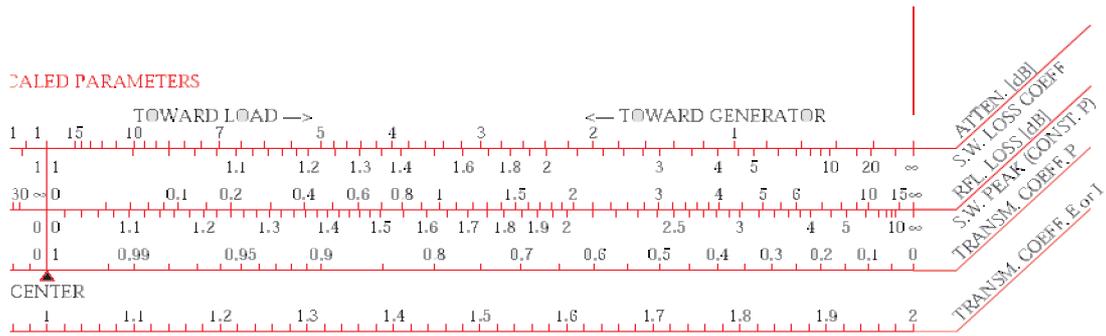}
\caption{Right part of the rulers depicted in Fig.\,\ref{ruler}}
\label{rulerright}
\end{figure}

A commonly used set of rulers that can be found below the Smith chart is depicted in Fig.\,\ref{ruler}. 
For further discussion, this ruler is split along the line marked centre, to a left (Fig.\,\ref{rulerleft}) and a right part (Fig.\,\ref{rulerright}) since they will be discussed separately for reasons of simplicity.
The upper part of the first ruler in Fig.\,\ref{rulerleft} is marked SWR which refers to the Voltage Standing Wave Ratio. The range of values is between one and infinity. One is for the matched case (centre of the Smith chart), infinity is for total reflection (boundary of the SC). The upper part is in linear scale, the lower part of this ruler is in dB, noted as dBS (dB referred to Standing Wave Ratio). Example: SWR = 10 corresponds to 20 dBS, SWR = 100 corresponds to 40 dBS (voltage ratios, not power ratios).

The second ruler upper part, marked as RTN.LOSS = return loss in dB. This indicates the amount of reflected wave expressed in dB. Thus, in the centre of the Smith chart nothing is reflected and the return loss is infinite. At the boundary we have full reflection, thus a return loss of 0 dB. The lower part of the scale denoted as RFL.COEFF. P = reflection coefficient in terms of POWER (proportional $|\Gamma|^{2}$). There is no reflected power for the matched case (centre of the Smith chart), and a (normalized) reflected power = 1 at the boundary.

The third ruler is marked as RFL.COEFF,E or I. With this, the modulus (= absolute value) of the reflection coefficient can be determined in linear scale. Note that since we have the modulus we can refer it both to voltage or current as we have omitted the sign, we just use the modulus. Obviously in the centre the reflection coefficient is zero, while at the boundary it is one. 

The fourth ruler is the voltage transmission coefficient. Note that the modulus of the voltage (and current) transmission coefficient has a range from zero, i.e., short circuit, to +2 (open = 1+$|\Gamma|$ with $|\Gamma|$=1). This ruler is only valid for $Z_{\text{load}}$ = real, i.e., the case of a step in characteristic impedance of the coaxial line.

The upper part of the first ruler in Fig.\,\ref{rulerright}, denoted as ATTEN. in dB assumes that an attenuator (that may be a lossy line) is measured which itself is terminated by an open or short circuit (full reflection). Thus the wave travels twice through the attenuator (forward and backward). The value of this attenuator can be between zero and some very high number corresponding to the matched case. The lower scale of this ruler displays the same situation just in terms of VSWR. Example: a 10\,dB attenuator attenuates the reflected wave by 20\,dB going forth and back and we get a reflection coefficient of $\Gamma$ = 0.1 (= 10\% in voltage).

The upper part of the second ruler, denoted as RFL.LOSS in dB refers to the reflection loss. This is the loss in the transmitted wave, not to be confused with the return loss referring to the reflected wave. It displays the relation $P_{\text{t}} = 1 - |\Gamma|^{2}$ in dB. Example: If $|\Gamma| = 1/\sqrt{2} =$ 0.707, the transmitted power is 50\% and thus the loss is 50\% = 3\,dB.

The third ruler (right), marked as TRANSM.COEFF.P refers to the transmitted power as a function of mismatch and displays essentially the relation $P_{\text{t}} = 1 - |\Gamma|^{2}$. Thus in the centre of the Smith chart there is a full match and all the power is transmitted. At the boundary there is total reflection and for a $\Gamma$ value of 0.5, for example, 75\% of the incident power is transmitted.
\end{document}